\documentclass[useAMS,usenatbib]{mn2e}
\usepackage{epsfig}
\usepackage{rotating}
\usepackage{color}
\usepackage{amsmath}
\usepackage{amssymb}
\usepackage{multirow}
\usepackage{multicol}
\usepackage{apjfonts}
\usepackage{url}
\usepackage{natbib}
\usepackage{bm}
\newcommand{\mr}[1]{\mathrm{#1}}

\newcommand{\Porb}{\ensuremath{P_{\rm orb}}}
\newcommand{\incl}{\ensuremath{i}}

\newcommand{\RSun}{\ensuremath{{\rm R}_{\odot}}}

\newcommand{\note}[1]{}
\newcommand{\omorb}{\ensuremath{\Omega_{\rm orb}}}
\title[Nonlinear mode coupling in KOI-54]{It takes a village to raise a tide: nonlinear multiple-mode coupling and mode identification in KOI-54}

\author[O'Leary \& Burkart]{Ryan M.\ O'Leary$^{1}$\thanks{Einstein Fellow}\thanks{E-mail: oleary@berkeley.edu} and  Joshua Burkart$^{1,2}$\\
$^{1}$Department of Astronomy and Theoretical Astrophysics Center, University of California, Berkeley, CA 94720, USA\\
$^2$Department of Physics, 366 LeConte Hall, University of California,
Berkeley, CA\ \ 94720, USA}

\begin{document}
\maketitle

\begin{abstract}
 We explore the tidal excitation of stellar modes in binary systems
 using {\em Kepler} observations of the remarkable eccentric binary
 KOI-54 (HD 187091; KIC 8112039), which displays strong ellipsoidal
 variation as well as a variety of linear and nonlinear pulsations. We
 report the amplitude and phase of over 120 harmonic and anharmonic
 pulsations in the system. We use pulsation phases to determine that
 the two largest-amplitude pulsations, the 90th and 91st harmonics,
 most likely correspond to axisymmetric $m=0$ modes in both stars, and
 thus cannot be responsible for resonance locks as had been recently
 proposed. We find evidence that the amplitude of at least one of
 these two pulsations is decreasing with a characteristic timescale of
 $\sim 100\,$yr.  We also use the pulsations' phases to confirm the
 onset of the traveling wave regime for harmonic pulsations with
 frequencies $\lesssim 50\; \omorb$, in agreement with theoretical
 expectations.  We present evidence that many pulsations that are not
 harmonics of the orbital frequency correspond to modes undergoing
 simultaneous nonlinear coupling to multiple linearly driven parent
 modes. Since coupling among multiple modes can lower the threshold
 for nonlinear interactions, nonlinear phenomena may be easier to
 observe in highly eccentric systems, where broader arrays of driving
 frequencies are available.  This may help to explain why the observed
 amplitudes of the linear pulsations are much smaller than the
 theoretical threshold for decay via three-mode coupling.
\end{abstract}


\section{Introduction}
\label{sec:intro}

Stars and planets in eccentric orbits exchange energy and angular
momentum through tidal interactions. The net tidal fluid response can
be conceptually divided into two components. The equilibrium tide is
the large-scale prolate distortion caused by nonresonantly excited
stellar modes \citep{1977A&A....57..383Z}. The dynamical tide, which
is our focus in this work, corresponds to low-frequency internal waves (gravity and inertial) that are
resonantly excited by the time-varying tidal potential.  Such waves
have much shorter damping times, and are expected to play an important
role in the circularization of orbits and spin up of stars
\citep[e.g.,][]{1975A&A....41..329Z,1989ApJ...342.1079G,2002A&A...386..222W}.
Although most prior work on tides has been calculated in the linear
regime, nonlinear effects may play an important role in redistributing
energy and angular momentum in binary systems on much shorter
timescales \citep[e.g.,][]{2010MNRAS.404.1849B,2012ApJ...751..136W}.

In this work, we explore the role of linear and nonlinear dynamical
tides in the recently discovered \emph{Kepler} system KOI-54 (HD
187091; KIC 8112039; \citealt{W11}, hereafter W11). In particular, we
aim to understand the nature of the largest-amplitude tidally excited
pulsations to ascertain if they are responsible for resonance locks, a
phenomenon that allows for efficient spin-orbit coupling in binary
systems \citep{2002A&A...386..222W}.  More generally, we explore the
harmonic and anharmonic pulsations that we observe in this particular
binary system, and address how their amplitudes and frequencies are
determined by a complex set of nonlinear interactions amongst many different modes.

Recently, \citet[][hereafter B12]{B12} developed a quantitative
framework for analyzing such pulsations, which they termed tidal
asteroseismology (see also \citealt{2011arXiv1107.4594F}). This work used theoretical
stellar models to investigate the linear, nonadiabatic response of
stars to linearly excited resonant oscillations.  This analysis
naturally explained many qualitative features of KOI-54's harmonic
pulsations. Nonetheless, several puzzles remained.

One unresolved question was the nature of the two most prominent
pulsations in the system, which have frequencies that are the 90th \&
91st harmonics of the orbital frequency. The amplitudes of the
pulsations ($294\,\mu$mag and $228\,\mu$mag, or parts-per-million, respectively) are considerably
larger than any of the other pulsations; indeed, B12 estimated that
the probability of their occurrence through purely linear excitation 
to be only  $\sim 1\%$.  Both
\citet[]{2011arXiv1107.4594F} and B12 considered the possibility that
the two pulsations could be occurring due to two $l=2$, $|m|=2$
g-modes responsible for resonance locks. \citet[]{2011arXiv1107.4594F} showed that
the frequencies of the 90th and 91st harmonics are within a few
percent of the natural, \emph{a priori} prediction of the most likely
frequency at which resonance locks should occur, given KOI-54's
observed system parameters. However, B12 found that the maximum torque
possible from likely modes was much too small to effect a resonance
lock, and also pointed out that it is unlikely for two resonance
locks to occur simultaneously.

B12 and \citet[]{2011arXiv1107.4594F} also reported that the two
largest anharmonic pulsations appear to be driven by the parametric
instability of the $91$st harmonic. However, B12 found that the
estimated amplitude for a mode to become overstable and drive the
parametric instability is $\sim 100\times$ larger than the observed amplitude of
the $91$st harmonic if it is an $m=0$ mode.

We address the nature of the harmonic pulsations by analyzing their
amplitudes and frequencies together with their phases.  B12 \citep[see
  also][]{2003A&A...397..973W} showed that the observed phase of the
pulsations depends on the excited mode's azimuthal order $m$, its
damping rate $\gamma$, and the difference $\delta \omega$ between the
eigenmode frequency and the driving frequency (see
eq.~\ref{eq:phase}).

In this work, we extend the work of W11 using additional publicly
available {\em Kepler} data to determine all of the observable properties of the
pulsations, in particular their phases, which were not originally
reported. This allows us to determine more information about the modes
responsible for the individual pulsations, in particular the azimuthal
order $m$.  Furthermore, with the
addition of five more quarters of data, and consequently greater
signal-to-noise ratios, we are able to search for pulsations with amplitudes lower than W11 were able to detect.

The structure of our paper is as follows.  We describe \emph{Kepler}
photometry and our data reduction routine in \S~\ref{sec:data}, and
give an overview of the observed pulsations in \S~\ref{sec:sin}. In
\S~\ref{sec:lin}, we analyze the tidally driven, linear oscillations
of KOI-54.  In \S~\ref{sec:nonlin}, we analyze the nonlinear
oscillations.  Finally, we summarize our results and
discuss their implications in \S~\ref{sec:disc}.

\begin{figure*}
\centering \includegraphics[width=\textwidth]{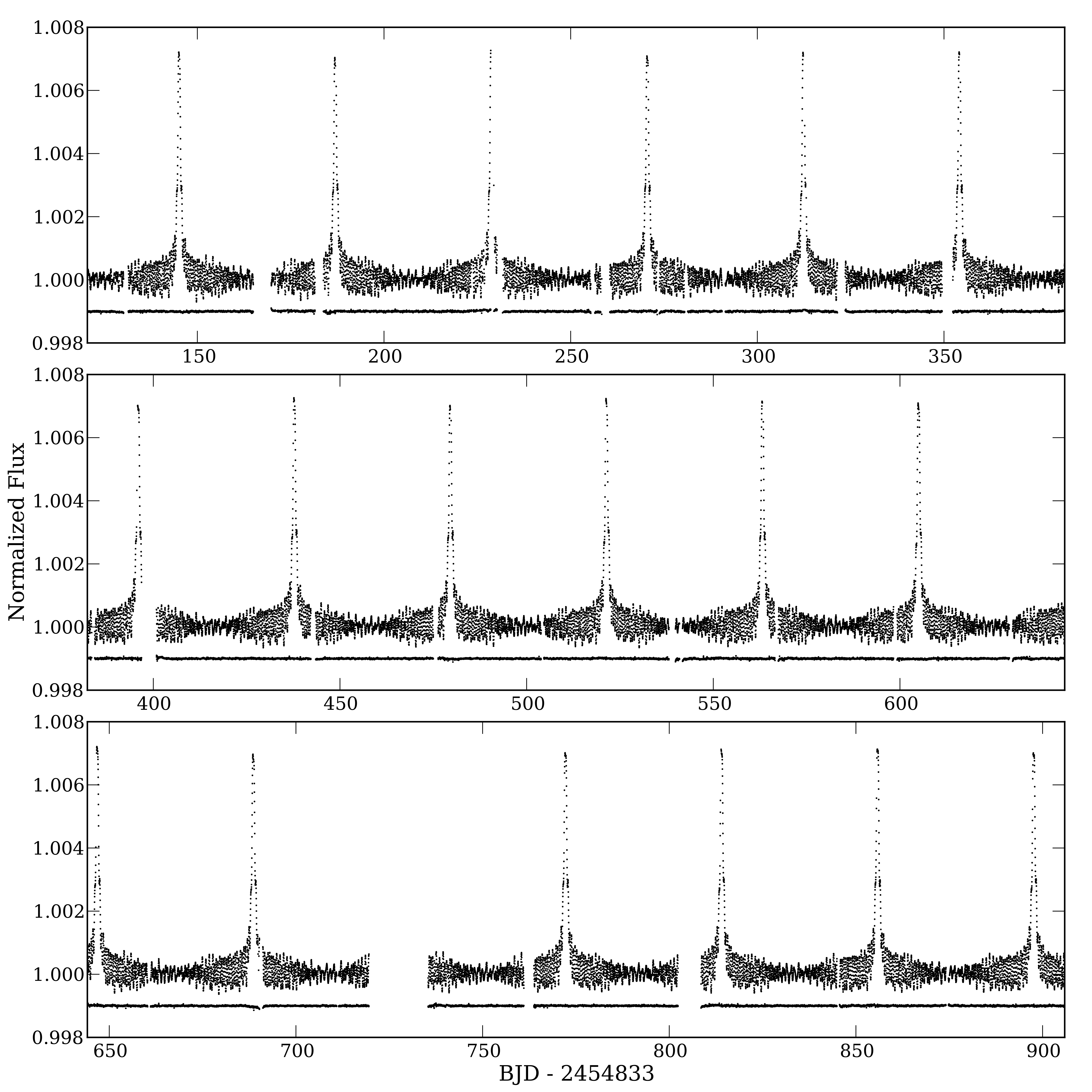}
\label{fig:reduced}
\caption{KOI-54 lightcurve.  Plotted is the normalized flux of KOI-54
  with the data detrended as in \S~\ref{sec:data}.  Below the
  lightcurve we show residuals from subtracting our
  ellipsoidal model and 182 of the best-fit sinusoidal
  pulsations.}
\end{figure*}

\section{\emph{Kepler} photometry and detrending}
\label{sec:data}
We use 785.4\,d of nearly continuous photometric observations of
KOI-54 that have been publicly released by the \emph{Kepler} mission
\citep{2010Sci...327..977B,2010ApJ...713L..79K}. During these
observations, the satellite underwent multiple safe-mode shutdowns, as
well as scheduled rollings that introduce large artifacts into the raw
data.  We detrend the data and remove these artifacts in the following
way.  We visually inspect the released raw data, and remove, by eye,
obvious outliers due to cosmic rays.  After inspection, we fit cubic
polynomials (and lines) to each series of contiguous data
simultaneously along with  a photometric model for the periastron
brightening events as well as 30 sinusoids for the largest-amplitude
pulsations. We then remove all outliers that deviate from the
remaining data by more than three standard deviations, and refit our
models and detrending curves. After we detrend the data, we
subsequently subtract the photometric model from the normalized
lightcurve and remove low-frequency pulsations with a Hann window
function of width $8\,$d.

W11 employed the proprietary {\sc ELC} modeling code
\citep{2000A&A...364..265O} to simultaneously fit the {\em Kepler}
photometry together with complementary radial velocity observations of
KOI-54. We instead use the much faster and simpler photometric model
detailed in B12 (Appendix B), which is sufficient to determine the
amplitudes, frequencies, and phases (and any variations) of the high-frequency ($f \gtrsim 20 \omorb$) dynamical tide with
minimal contamination from the equilibrium tide and stellar
irradiation.  This linear model decomposes the stellar flux
perturbation induced by the equilibrium tide into spherical harmonics,
making use of von Zeipel's theorem \citep{1924MNRAS..84..665V}. The
effects of irradiation of each star by its companion are also
included, and are modeled as absorption and immediate reemission at the photosphere. We evaluate changes in
luminosity due to the equilibrium tide and irradiation up to spherical
harmonic degree $l=3$; higher-degree harmonics do not contribute
significantly to the signal.

To model limb darkening, we use the four-coefficient nonlinear fit
found in \citet{2011MNRAS.413.1515H} determined for the \emph{Kepler}
bandpass for an $8,800 $\,K star with a surface gravity $\log g = 4$,
consistent with best-fit parameters of both stars in W11. We then find
the least-squares best fit for the orbital period, epoch of
periastron, and bandpass correction coefficients $\beta_1$ \&
$\beta_2$. For the remaining parameters, we use the best-fit values in
W11.  In principle, it should be possible to determine $\beta_1$ and
$\beta_2$ \emph{a priori}, as was done in W11. We find that for the
objectives of this paper, by fitting for $\beta_1$ and $\beta_2$ and
subtracting the ellipsoidal variation and irradiation from the
lightcurve, we do not introduce any significant errors in our
assessment of the high-frequency pulsations.  The parameters of KOI-54
used in this study are listed in Table~\ref{tab:params}.

\begin{table}

\caption{System parameters of KOI-54. From top to bottom are the derived properties of the binary from spectroscopic observations, best-fit system parameters from the photometric model of W11,
  best-fit parameters using our photometric model, and the tabulated coefficients used in the photometric model in this work.\label{tab:params}}
 \centering
 \begin{tabular}{llrrc}\cline{2-5}
 & parameter & value& error & unit\\\cline{2-5}
 \multicolumn{3} { |l| }{ Observations (W11)}\\

 \multirow{7}{*}& $T_{1}$     & 8500  & 200  & K \\
 &$T_{2}$     & 8800  & 200  & K \\
 &$L_2/L_1$ & 1.22 & 0.04 & \\ 
 &$v_{\mr{rot},1} \sin{i_1}$ & 7.5  & 4.5 & $\mr{km}/\mr{s}$ \\
 &$v_{\mr{rot},2} \sin{i_2}$ & 7.5  & 4.5 & $\mr{km}/\mr{s}$ \\
 &$[\text{Fe}/\text{H}]_1$  & 0.4  & 0.2  &   \\
 &$[\text{Fe}/\text{H}]_2$  & 0.4  & 0.2  &   \\
 \cline{2-5}
 \multicolumn{3} { |l| }{ Lightcurve modeling (W11)}\\
 \multirow{11}{*}&  $\Porb$                 & 41.8050  & 0.0003  & days \\%
 &$T_{\rm p}$                 & 2455103.5490  & 0.0010  & BJD \\%
 &$e$           & 0.8335  & 0.0005  &      \\
 &$\omega$         & 36.70   & 0.90    & degrees \\
 &$\incl$            & 5.50     & 0.10    & degrees \\
 &$a$                 & 0.3956    & 0.008   & AU    \\
 &$M_{2}/M_{1}$  &        1.024   & 0.013 &\\
 &$R_{1}$        & 2.20   & 0.03  & $\RSun$ \\
 &$R_{2}$        & 2.33   & 0.03  & $\RSun$\\ \cline{2-5}
 \multicolumn{3} { |l| }{ Lightcurve modeling (this work)}\\

\multirow{6}{*} & $\Porb$ & 41.8050 & 0.0001 & days\\
&$T_{\rm p}$ & 2455061.73814 & 0.006 & BJD\\

&$\beta(T_1)$ & 0.489 & & \\
&$\beta(T_2)$ & 0.818 & &\\
\cline{2-5}
\multicolumn{3} { |l| }{ Limb darkening coefficients (Howarth 2011)}\\

\multirow{6}{*} &  $b_0$ & 1.0000 & & \\
&$b_1$ & 0.7076& &\\
&$b_2$ & 0.3230& &\\
&$b_3$ & 0.0596& &\\
&$c_0$ & 0.0000& &\\
&$c_1$ & 1.4152& &\\
&$c_2$ & 1.9380& &\\
&$c_3$ & 0.7154& &\\\cline{2-5}

 \end{tabular}\rule{20pt}{0pt}
\end{table}

\begin{figure}
\centering \includegraphics[width=\columnwidth]{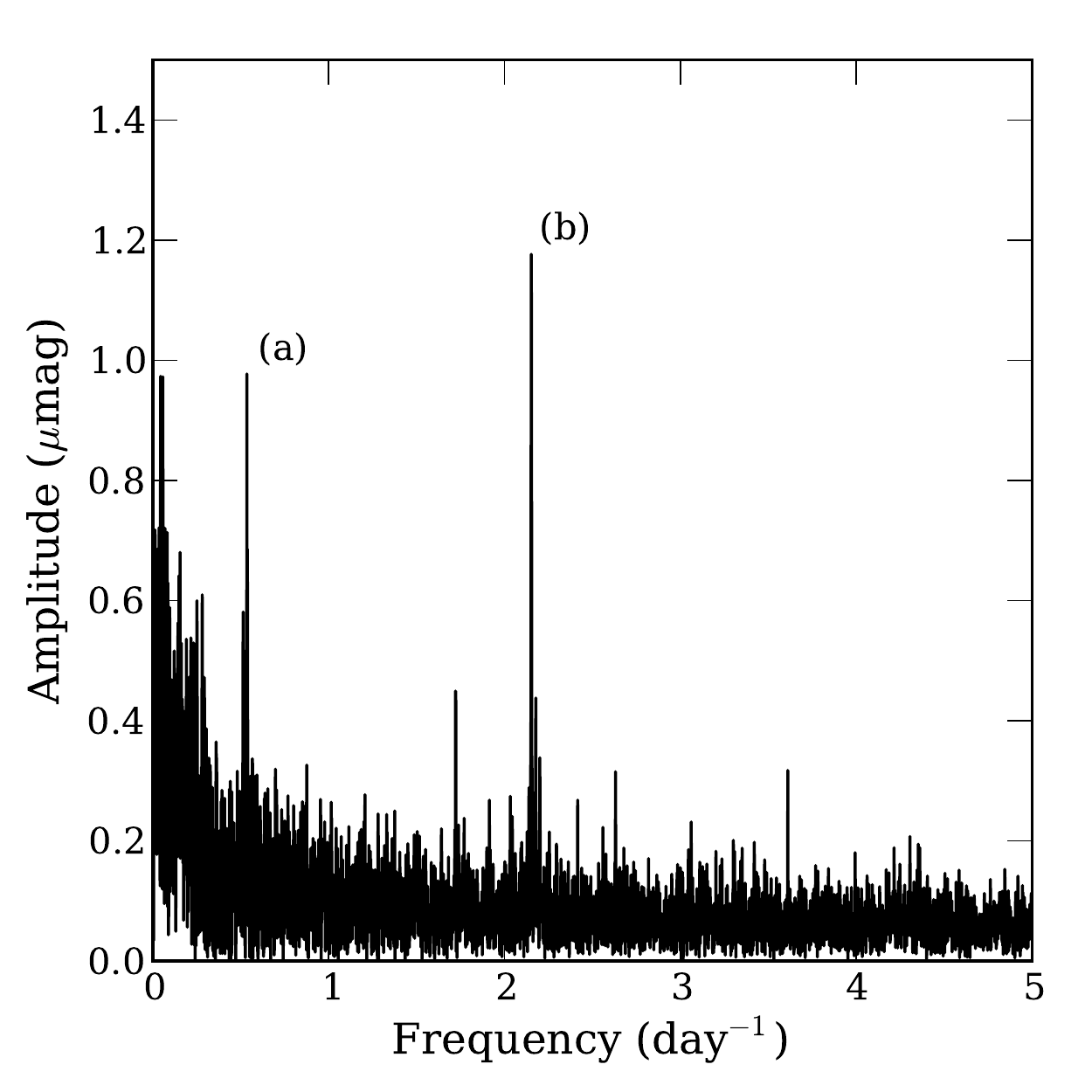}
\caption{Frequency spectrum of residual data after removing the ellipsoidal variation as well as 182 of the largest-amplitude pulsations (see~\S~\ref{sec:osc}).  No pulsations have amplitude $\gtrsim 1.2 \mu$mag. In the figure we have marked two of the largest-amplitude pulsations that we did not include in our global fits, as they have frequencies that are too close to other frequencies to be independently resolved in the data. The frequency marked (a) is close to F5, the largest-amplitude anharmonic pulsation. The peak marked (b) is close to F1, the largest-amplitude harmonic pulsation $f = 90 \omorb$. }
\label{fig:spectrum}
\end{figure}

Compared to W11, who detrended the data by completely masking the
brightening events and then similarly fitting cubic polynomials, our
method gives smaller residuals near the brightening events and smaller
formal errors in the derived properties of the pulsations.
 
\section{Sinusoidal pulsations in KOI-54}
\label{sec:sin}
\subsection{Observed frequencies, amplitudes, and phases}
\label{sec:osc}

We systematically search for the $200$ largest-amplitude\footnote{In
  practice our search routine looks for the pulsations with the
  most power in a fast \citep{1989ApJ...338..277P} Lomb--Scargle
  periodogram \citep{1976Ap&SS..39..447L,1982ApJ...263..835S}.}
pulsations in the lightcurve by removing our best-fit ellipsoidal
variation model and then iteratively removing the largest-amplitude
pulsation found in the power spectrum of the residual data.  Between
steps, we simultaneously refit the data for the frequency, phase,
and amplitude of all known pulsations.  In practice, the process of
refitting the data via least-squares can cause the frequencies of two
or more initially distinct pulsations to converge.
As a result, we remove 18 cases where the frequency spacing between two pulsations
is less than the reciprocal of the total duration of observations (i.e., pairs that have not undergone one full beat in the window of observations), and simultaneously refit all the remaining $182$ pulsations.

 In Table~\ref{tab:fitsh}, we present the $70$ largest-amplitude {\em
   harmonic}  pulsations with
 frequency $f >20\omorb$ and $|f - k \omorb| < 0.02\omorb$, where $k$
 is an integer,\footnote{We choose this specific value of the frequency
   offset because no observed pulsations were observed with an offset
   between $0.019\omorb$ and $0.035\omorb$. } along with their measured phase (from 0 to 1) with
 respect to the epoch of periastron as well as BJD$-2454833$. In
 Table~\ref{tab:fitsa}, we similarly present the $59$
 largest-amplitude {\em anharmonic} pulsations with $|f - k \omorb|
 \ge 0.02\omorb$.  The derived properties of low-frequency pulsations
 ($f < 20\omorb$) have much greater uncertainties because of
 systematic trends in the raw data that we were unable to remove
 completely, so we do not report these pulsations in our results.
 After removing the $182$ largest-amplitude pulsations, we are left
 with no pulsations with amplitude $\gtrsim 1.2\,\mu$mag. To be
 consistent with W11 and much of the observing literature, we fit for
 the amplitude $A_i > 0$, frequency $f_i$ and phase offset $\delta_i$,
 using the form $A_i \sin{2\pi (f_it+\delta_i)}$, where the time $t$
 is measured relative to the epoch of periastron, $t_p$, as well as
 BJD-$2454833$.  The variance in the residual lightcurve is $\approx
 16\, \mu$mag, approximately 1.6 times larger than the photometric
 uncertainty of each observation.  The frequency spectrum of the
 residuals is presented in Figure~\ref{fig:spectrum}.

We estimate the formal uncertainties of our fits by a Monte Carlo
method (for systematic uncertainties, see \S~\ref{sec:unc}).  We
create mock light curves using the best-fit parameters of the
pulsations, using the same observing window as the real data. We then
add Gaussian white noise to each point with an amplitude $10^{-5}$,
consistent with the photometric uncertainty in each data
point. Finally, we add $1/f$ noise to the data by passing it through
the same high-pass Hann filter as the data, and then rescaling the
amplitude of the $1/f$ noise so the final noise amplitude of the mock
light curve equals the observed amplitude of the residuals. The power
spectrum of the noise we generate matches both the amplitude and shape
of the residual power spectrum much better than white noise alone,
which is commonly used in other analyses.  We refit each mock data set
100 times using the previously known best-fit parameters as our
initial values, and report the variance of the best-fit parameters as
the formal uncertainties in Tables~\ref{tab:fitsh} \&
\ref{tab:fitsa}. 

For some pulsations that have similar frequencies, we find that
fitting routine can be pathological, resulting in the two pulsations
to be fit with the same frequency.  Even when this happens in only one
of the noise realizations, the reported formal uncertainty can be
significantly larger than the intrinsic amplitude of the observed
pulsation. In Tables~\ref{tab:fitsh} \& \ref{tab:fitsa} we report all
observed pulsations with amplitudes larger than $\approx 0.7 \mu$mag,
which typically are $\approx 3\sigma$ local detections, and have not
been corrected for searching over the entire parameter space.  Some of
these pulsations, especially the low-frequency anharmonic pulsations,
may be noise. We choose to report all of these pulsations because any
spurious pulsations do not contaminate our analysis.

\subsection{Systematic uncertainties}
\label{sec:unc}

\begin{figure*}
\centering \includegraphics[width=\columnwidth]{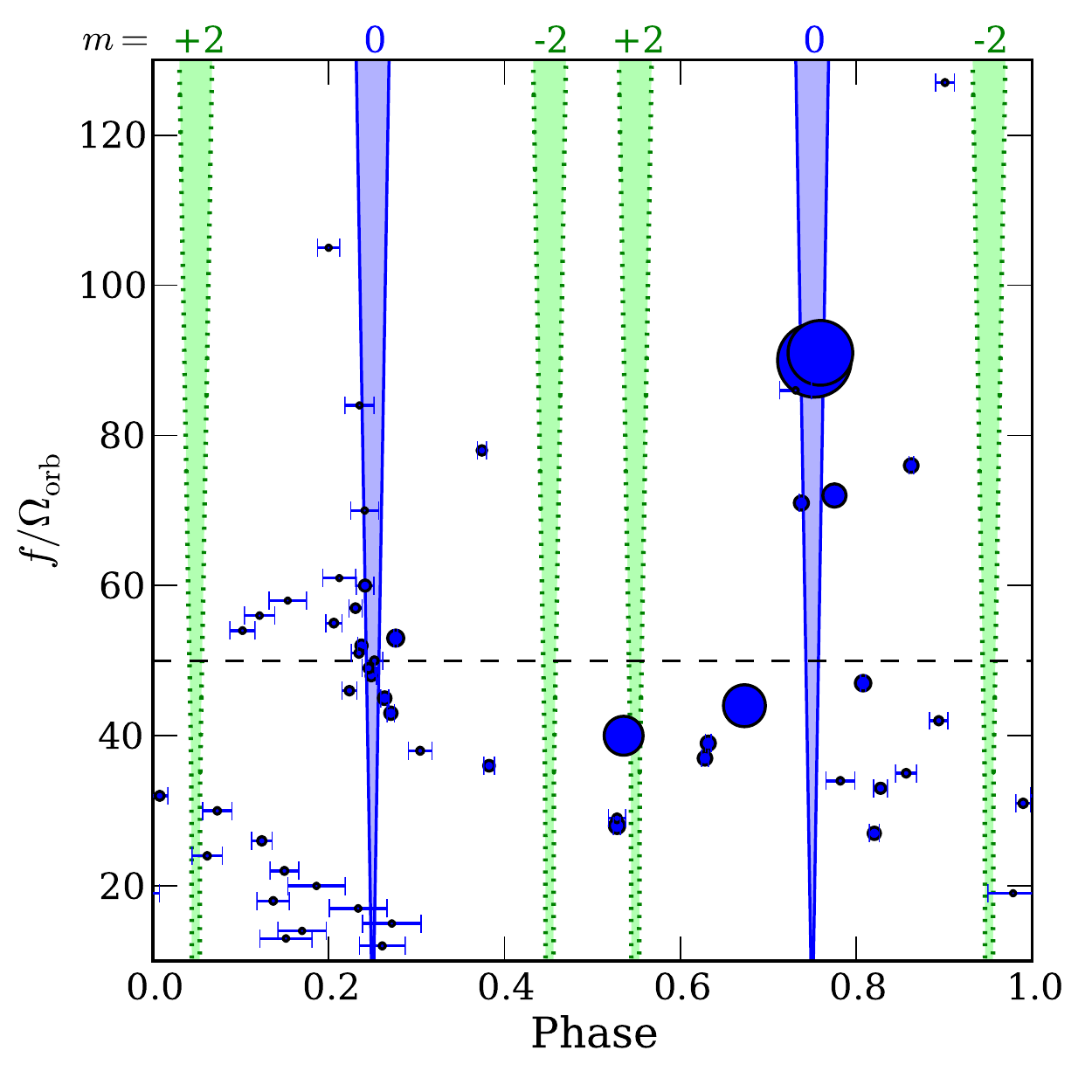}
\includegraphics[width=\columnwidth]{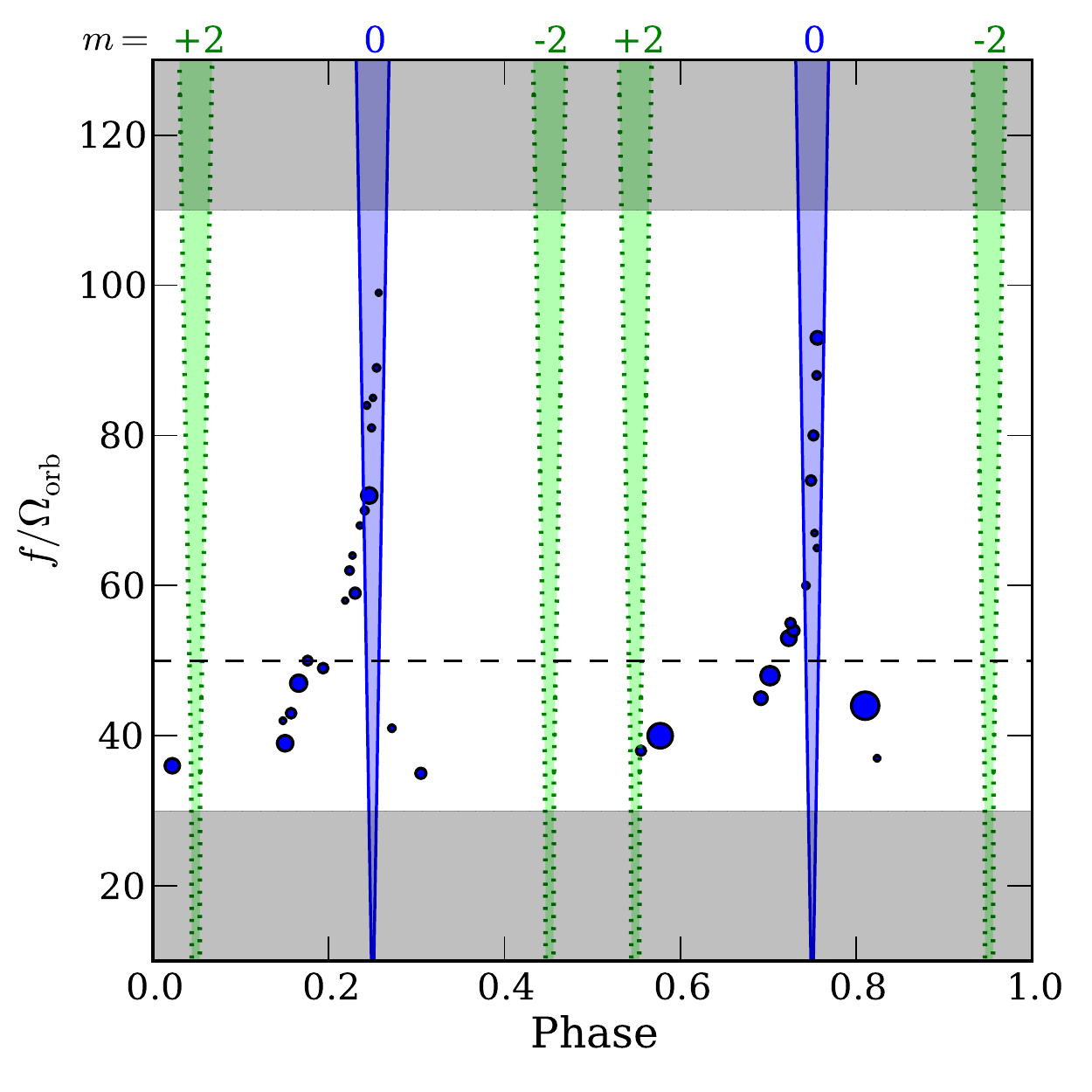}

\caption{Phases of harmonic pulsations. {\em Left:} Plotted are phases
  of the near-perfect harmonics ($|f - k \omorb| < 0.02\omorb$) of the
  orbital frequency relative to the best-fit epoch of periastron found
  by W11.  The area of each circle corresponds to the amplitude of the
  pulsation. The vertical shaded cones correspond to the expected
  phase for $m=0$ (blue with solid lines) and $m=\pm 2$ (green with
  dotted lines) modes.  The width of the cones result from the
  systematic uncertainties in the epoch of periastron as well as the
  argument of periastron.  We conservatively assume that the
  systematic uncertainty in the epoch of periastron is $6\times$
  larger than the quoted uncertainty in W11, consistent with our
  modeling.  The actual system parameters, of course, correspond to
  only one line within  each cone.  For oscillations that are in a
  resonance lock, the phase of the pulsation can be nearly arbitrary.
  We see the 90th and 91st harmonics are consistent with being $m = 0$
  oscillations.  For harmonics with $f/\omorb \lesssim 50$, the
  oscillations are expected to be traveling waves, and may have an
  arbitrary phase. We omit harmonics that have phase uncertainties
  greater than $0.015$. {\em Right:} The phase information of linearly
  driven oscillations in a theoretical binary system from B12, which
  is qualitatively similar to KOI-54.  Only the response of $l=2$,
  $m=0$ oscillations are included in the calculation.  We do not show
  results in the shaded regions where the linear response was not
  calculated. There is no uncertainty in the epoch of periastron in
  the theoretical model; the expected phases for low-amplitude $l=2$,
  $m=0$ pulsations should be offset from the center of the shaded
  cones.}\label{fig:scatterphase}
\end{figure*}

Our work here, and work on similar systems, inevitably possesses
important systematic errors. The greatest uncertainty in our
calculations is the systematic uncertainty in fitting for the epoch of
periastron, which impacts the derived phase offsets of the harmonic
pulsations.  Indeed, in W11, the authors' best-fit epoch of periastron
derived from radial velocity data alone is $6\sigma$ away from the
epoch found using radial velocities together with photometry.
Visually, the systematic errors in the fitting of the RV data appear
more significant.  The quoted uncertainties in the W11 RV data were
also probably too low due to heterogeneity of the spectra (G. Marcy,
private communication).  For this reason, we exclusively use the
photometrically determined epoch of periastron.  In addition, fitting
only $15$ of the largest-amplitude pulsations, as done in W11, is
likely insufficient for determining the epoch of periastron.  As we
will show in \S~\ref{sec:lin} (see Fig.~\ref{fig:scatterphase}), many
of the low-amplitude pulsations are $m=0$ oscillations that reach
their maximum near periastron and bias the fitting of the ellipsoidal
variation and irradiation of the binary.  Indeed we found that the
peak amplitudes of the observed brightening events were $80\,\mu$mag
larger than the best-fit model if the pulsations are not taken into
consideration.

In W11, the authors compensated for the undetected pulsations by
increasing the uncertainty of each data point by a factor of $10$, thereby
obtaining a normalized $\chi^2$ less than unity. Nevertheless,
systematic errors from the undetected pulsations may remain. We attempt to
ascertain how the pulsations bias the epoch of periastron by
refitting the detrended data simultaneously with 15 or 30 of the
largest-amplitude pulsations.  We find systematic changes in the epoch
of periastron approximately three times larger than the uncertainty
reported in W11. We conservatively estimate that the true uncertainty
is six times larger than reported in W11 (Table 2).

 Unfortunately, many of the low-frequency pulsations are systematic
 errors from our fitting routine using cubic polynomials. We therefore
 fit the pulsations by both including all low-frequency pulsations and
 compare this to the fit excluding pulsations with $f < 20 \omorb
 \approx 0.5\,$day$^{-1}$.  We find that the difference in the fits
 of the high-frequency pulsations are less than the formal
 uncertainties listed in Tables~\ref{tab:fitsh} \& \ref{tab:fitsa}.
 We also note that we do not include the impact of Doppler boosting
 in our light curve models.  We find that including the impact of
 boosting does not change the amplitude of any pulsations by more than
 $0.5\,\mu$mag, although the Doppler boosting signal is important in other
 eccentric binaries reported by {\em Kepler} that show strong tidal
 distortions \citep{2012ApJ...753...86T,2013arXiv1306.1819H}.

Incompletely subtracting the ellipsoidal variation can also
systematically contaminate our fits of the harmonic pulsations.
Fortunately, the Fourier decomposition of the ellipsoidal variation is
continuous with frequency. We estimate that the total contamination is
$\lesssim 0.3\mu$mag by inspecting the limits we placed on the
amplitude of undetected harmonic pulsations. A similar technique may
be useful when constraining the contamination of lower harmonics in
less eccentric binaries.

\section{Linearly driven harmonic pulsations}
\label{sec:lin}

Each star is expected to have tidally driven pulsations at perfect
harmonics of the orbital frequency $\omorb$.  In \S~\ref{sec:linintro}
we briefly outline the general theory of dynamical tides and describe
how the pulsations reveal detailed properties of the star (B12,
\citealt[]{2011arXiv1107.4594F}).  In \S~\ref{sec:harm} we analyze the
phases of the harmonic pulsations relative to periastron in order to
determine the azimuthal order, $m$, of the oscillations. In
\S~\ref{sec:time}, we search for time variability in the amplitudes, phases, and
frequencies of KOI-54's pulsations.

\subsection{Introduction}
\label{sec:linintro}
For a nonrotating star, the time-varying tidal potential excites $l
\geq 2$ mode spherical harmonics in the star. The oscillations within
the star are observed as sinusoidal pulsations,\footnote{We distinguish
  the intrinsic changes within the star, i.e., {\em oscillations},
  from the extrinsically observed {\em pulsations}.} which are
averaged over the disk of the entire star.  The steady-state,
equilibrium solution is simply a sinusoidal pulsation with an observed
frequency $k\omorb$ that is a harmonic of the orbital frequency, where $k$ is
an integer. The amplitude and phase of the pulsation are determined
by how well tuned the oscillator is relative to an eigenfrequency as measured by the detuning $\delta \omega = \omega_{nl} -
k\omorb$, by the damping rate of the excited mode $\gamma_{nl}$, as well
as by the spatial coupling between the driving force and the mode \citep{1977ApJ...213..183P}.  By
directly measuring the amplitude and phase of a harmonic pulsation, it
is possible to determine the detuning-to-damping ratio $\delta \omega/
\gamma_{nl}$, as well as the azimuthal order of the oscillation, $m$.

The phases of observed pulsations of a star are determined by disk
averages of the flux perturbations on the stellar surface. In this work we measure the phase of the pulsation from the light curve using the equation (see \S~\ref{sec:osc})
\begin{equation}
\label{eq:fit}
A_i \sin{2\pi (f_it+\delta_i)},
\end{equation}
where $A_i >0$, and $t$ is measured relative to the epoch of
periastron, $t_p$.  The observed phase of the pulsation directly
depends on the order of the harmonic, $m$.  B12 derived the observed
pulsation phase relative to the epoch of
periastron to be\footnote{This equation is derived from Eq.~33 of B12.
  B12 incorrectly included a $\pm$ on the right hand side of Eq.~33,
  which we have omitted here.  In addition, B12 defined the phase in
  radians using the cosine function.}
\begin{equation}
\label{eq:phase}
\delta = \left(\frac{1}{4}+\Psi_{nlmk} + m \phi_0 \right) \mod \frac{1}{2},
\end{equation}
where $\phi_0 = 1/4 - \omega_p/(2\pi) \mod 1$, $\omega_p$ is the
argument of periastron of the orbit,  
\begin{equation}
\Psi_{nlmk} = -\frac{1}{2\pi}
\arctan{\left(\frac{\omega_{nl}^2-\sigma_{km}^2}{2 \gamma_{nl}\sigma_{km}}\right)} \mod  \frac{1}{2},
\label{eq:PSI}
\end{equation}
 and $\sigma_{km} = k\omorb - m \Omega_*$ is the driving frequency of
 the tide in the corotating frame of the star with rotation frequency
 $\Omega_*$.  We expect that most excited oscillations are poorly
 tuned, i.e., $|\delta \omega| \gg \gamma_{nl}$, since the frequency spacing
 of the eigenmodes is much larger than the damping rate of 
 high-frequency g-modes.  In this limit, equations \ref{eq:phase} \&
 \ref{eq:PSI} reduce to
\begin{equation}
\label{eq:finalphase}
\delta = \left(\frac{1}{4} + m \phi_0\right) \mod \frac{1}{2}.
\end{equation}
Since $\phi_0$ is a quantity derived from modeling the equilibrium
tide as well as the radial velocity data, we can directly determine
the order of an oscillation $m$ in this limit. 

In this work, we report the absolute phases of the pulsations $\in
[0,1)$ relative to the epoch of periastron using the sine function to
  be consistent with the observing literature (see
  \S~\ref{sec:osc}). A pulsation is observed at
  its maximum at the epoch of periastron when $\delta = 1/4$.  For
  $m=0$ modes, based on equation~\ref{eq:finalphase} we would expect the pulsation to be near maximum
  ($\delta \approx 1/4$) or near minimum ($\delta \approx 3/4$) at the
  epoch of periastron. B12 show that if the largest-amplitude
  harmonics of KOI-54 are $m=0$ modes, then their detuning-to-damping
  ratio is $|\delta \omega|/\gamma_{nl} \sim 20$ and so should be
  offset from $1/4$ or $3/4$ by no more than a few percent. Given the
  face-on orientation of KOI-54, most observable pulsations should be
  $m=0$, $l=2$ modes. Large amplitude $m=2$, $l=2$ pulsations may also
  be present, but the amplitudes of these pulsations are suppressed by
  a factor of $\sin^{|m|}{i} \approx 1/200$.

There are two other regimes where equation~\ref{eq:finalphase} does not
apply. First, when  oscillations are no longer  standing waves but are instead in the traveling regime, the phases begin to deviate from equation~\ref{eq:finalphase}, which we discuss in
more detail in \S~\ref{sec:travel}. Second, when an excited mode is in
a resonance lock, it is possible that $\delta
\omega / \gamma_{nl} \sim 1$, allowing for a potentially arbitrary phase. Such a resonance lock can only occur when $m \neq
0$. For stars similar to KOI-54, each star is expected to be in a
resonance lock only $\sim 10\,$percent of the
time, and as discussed in B12, it is much rarer for both stars to be
in a simultaneous lock.

As we have discussed, for most of the large-amplitude pulsations, only
one eigenmode of a single star will contribute to the corresponding
pulsations.  However, \emph{Kepler} photometry is so precise that we
are able to detect pulsations with amplitudes comparable to $1\,\mu$mag. A fraction of these low-amplitude pulsations will have
contributions from both stars or from multiple modes in a single star.

\begin{figure*}
\centering \includegraphics[width=\textwidth]{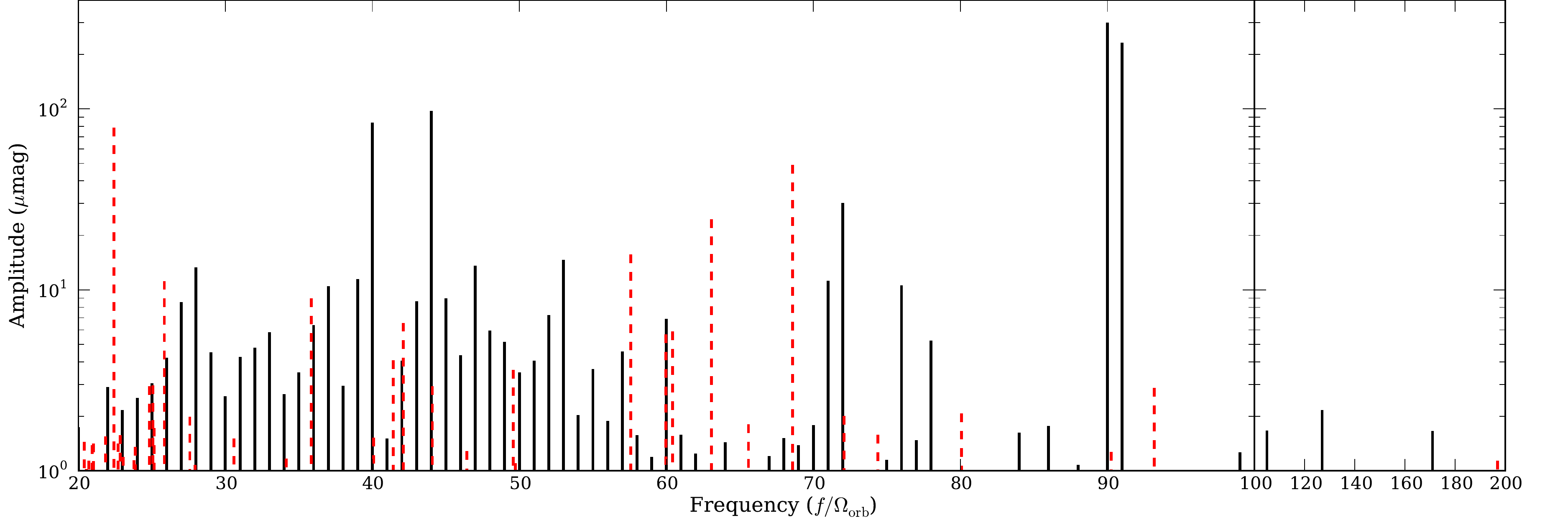}

\caption{Distribution of pulsation amplitudes down to
  $1\,\mu$mag.  The pulsations with frequencies consistent with
  perfect harmonics of the orbital frequency ($|f - k\omorb| < 0.02$) are shown with black
  lines.  The anharmonic pulsations are shown with red dashed lines. Note: the frequency scale changes at $f/\omorb = 100$. }
  \label{fig:spec} 
\end{figure*}

\subsection{Harmonic pulsation phases}
\label{sec:harm}
\subsubsection{Standing waves}

In Figure~\ref{fig:scatterphase}, we plot the phases of the
harmonic pulsations as well as the phases in a theoretical
model from B12.  The model was derived to qualitatively match the
amplitudes and frequencies of observed pulsations in KOI-54.  In both
plots, the area of each point is in proportion to the observed
amplitude of the pulsation, the error bars show the uncertainties in
the phase, and the filled vertical regions of the plot show the phases
expected for $m=0$ and $m=\pm2$ modes (\S~\ref{sec:linintro},
equation~\ref{eq:finalphase}).  For standing waves, with $k \equiv
f/\omorb \gtrsim 50$ (see \S~\ref{sec:travel} for traveling waves), we
expect the $m=0$ oscillations to have phases near $1/4$ and $3/4$
(\S~\ref{sec:linintro}).  Many of the observed pulsations are
consistent with this result (e.g., F1, F2, F7, F11, and F16 among
others).  In particular we find that the phases of the largest-amplitude 90th and 91st
harmonics, $.752 \pm .0001\,$(statistical)$\,\pm 0.013\,$(systematic)
and $.759 \pm .0001\,$(statistical)$\,\pm
0.013\,$(systematic\footnote{The systematic uncertainty includes the
  uncertainty in the epoch of periastron, and is proportional to the
  pulsation frequency.}) are consistent with being $m=0$ modes.  If
either of these pulsations were in a resonance lock, its phase could 
be arbitrary (when $\delta \omega / \gamma_{nl} \sim 1$;
\S~\ref{sec:linintro}).

There are a number of pulsations that appear to be $m \ne 0$ modes, e.g., F18, F32, F67, \& F68.  Not
shown in the figure is the pulsation with $k = 171$ (F68) which has a
phase near $0.5$. This is consistent with what we expect for a
typical $m=2$ oscillation, especially given its low observed
amplitude, which excludes it as originating from a resonance lock.  The phase of
the 105th harmonic (F67) is closer to the phase expected for an
$|m| = 3$ oscillation.  The phase of the 76th (F18) and
127th (F57) harmonics are consistent with $m=1$ modes. 

There are at least four possible explanations for why some pulsations
do not have phases that correspond to $m=0$ modes in
Figure~\ref{fig:scatterphase}. 1) The oscillation is an $m\ne 0$,
$l\ge2$ mode. 2) Given the large number of excited pulsations, we
might expect a few of the pulsations from each star to interfere. This
causes the observed phase to shift between the intrinsic phase of each
star when $\delta_A \ne \delta_B$. The observed phase of the pulsation, $\delta_{AB}$, is 
\begin{equation}
\delta_{AB} =  \frac{1}{2\pi} \cos^{-1}{\left(\frac{A_A \cos{2 \pi \delta_A}+A_B\cos{2 \pi \delta_B}}{A_{AB}}\right)}
\end{equation}
where the amplitude of the resulting pulsation is
\begin{equation}
A_{AB} = \sqrt{A_A^2+A_B^2+2A_AA_B\cos{2\pi(\delta_A-\delta_B})}.
\end{equation}
  3) If the oscillation is primarily driven by nonlinear terms via
  three or higher mode coupling its phase will no longer reflect the
  expected linear offset (see~\S~\ref{sec:nonlin}).  4) Lastly, if it
  is an $m \ne 0$ oscillation that was shifted from the traveling wave
  regime because of the star's rotation, then it could also have an
  arbitrary phase.  For our outliers, this is possible only if
  $\Omega_* \gtrsim 13 \omorb$, which is satisfied by the expected
  rotation rate of the star (B12).

It is unlikely that 90th and 91st harmonic pulsations originate in the
same star. For $k\sim 90$, the typical frequency spacing between the
eigenfrequencies of $m=0$, $l=2$ g-modes is significantly larger than
$\Delta k \sim 1$.  Indeed, the g-mode eigenfrequencies of a star are
expected to behave asymptotically as \citep{oscillations}
\begin{equation}
\label{eq:asymp}
\omega_{ln} \approx \omega_0 \frac{l+1/2}{(n+\alpha)\pi},
\end{equation}
where we assume that $l=2$ and $n \gg 1$.  Assuming that $m=0$ for
both the $k=90$ and $k=91$ pulsations,
we can estimate where other large amplitude $m=0$ fluctuations should
exist using stellar models or by tuning equation~\ref{eq:asymp}.  Just
as importantly, when a pulsation appears with the frequency between
two consecutive eigenmodes, we can conclude $|m| > 0$, since it must
have been shifted from stellar rotation by $m\Omega_*$. We can
calibrate equation~\ref{eq:asymp} by finding the eigenfrequencies of
stellar models with parameters near the best-fit found by W11. We use
an untuned MESA stellar model \citep{2011ApJS..192....3P} for a star
similar to those in KOI-54 to calibrate equation~\ref{eq:asymp} at
$n\approx 10$--$20$. We find the best-fit parameters are $\alpha
\approx 1.07$ and $\omega_0 = 40.5\,$d$^{-1}$.

Assuming that $n=14$ for $k = 90$, we roughly estimate that the next few
eigenfrequencies are near $83.5$, $79$, $74.5$, $70.6$, $63.8$,
$60.8$, $58.1 \omorb$, where the uncertainty is of order $\Delta k
\approx 1$.  Interestingly there are two large-amplitude pulsations in
KOI-54 with $k=71$ and $k=72$, both with phases consistent with $m=0$,
which may correspond to the $n=18$ modes of each star. There are no
pulsations with amplitudes $\gtrsim 1.7\,\mu$mag between where we
expect the $n=14$ and $n=15$ pulsations. Identifying the
eigenfrequencies of the stars in this manner can greatly reduce the
large computational overhead of finding stellar models that reproduce
the properties of the two stars.  The frequency spacing between the
high-frequency pulsations, with $f \gtrsim 100\omorb$, is also
consistent with the MESA models of B12.

  Figure~\ref{fig:scatterphase} shows how the
phases of the pulsations are a powerful tool to determine the nature
of the pulsations (e.g., the azimuthal order, $m$, or whether a mode is in a
resonance) as well as to verify the epoch of periastron and $\phi_0$.

\subsubsection{Traveling waves}
\label{sec:travel}

\begin{figure}
\centering \includegraphics[width=\columnwidth]{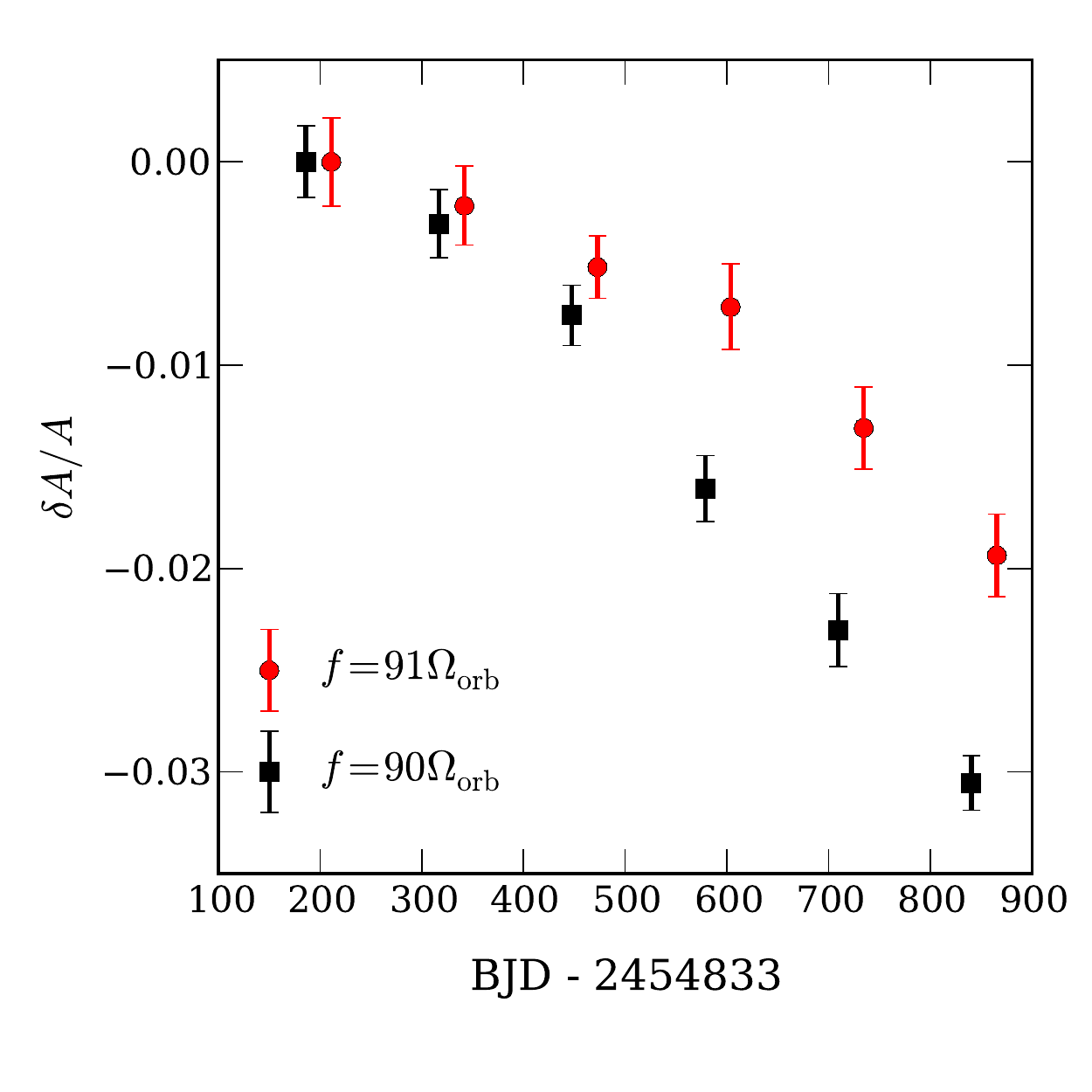}
\caption{Fractional amplitude change ($\delta A/A$) of the $k = 90$
  and $k = 91$ pulsations over time. The error bars show the estimated
  standard deviation of the fits for each individual time bin. The $k=91$ data points are offset from the center of the bin in order to increase legibility.}
\label{fig:ampchange}
\end{figure}

\begin{figure}
\centering \includegraphics[width=\columnwidth]{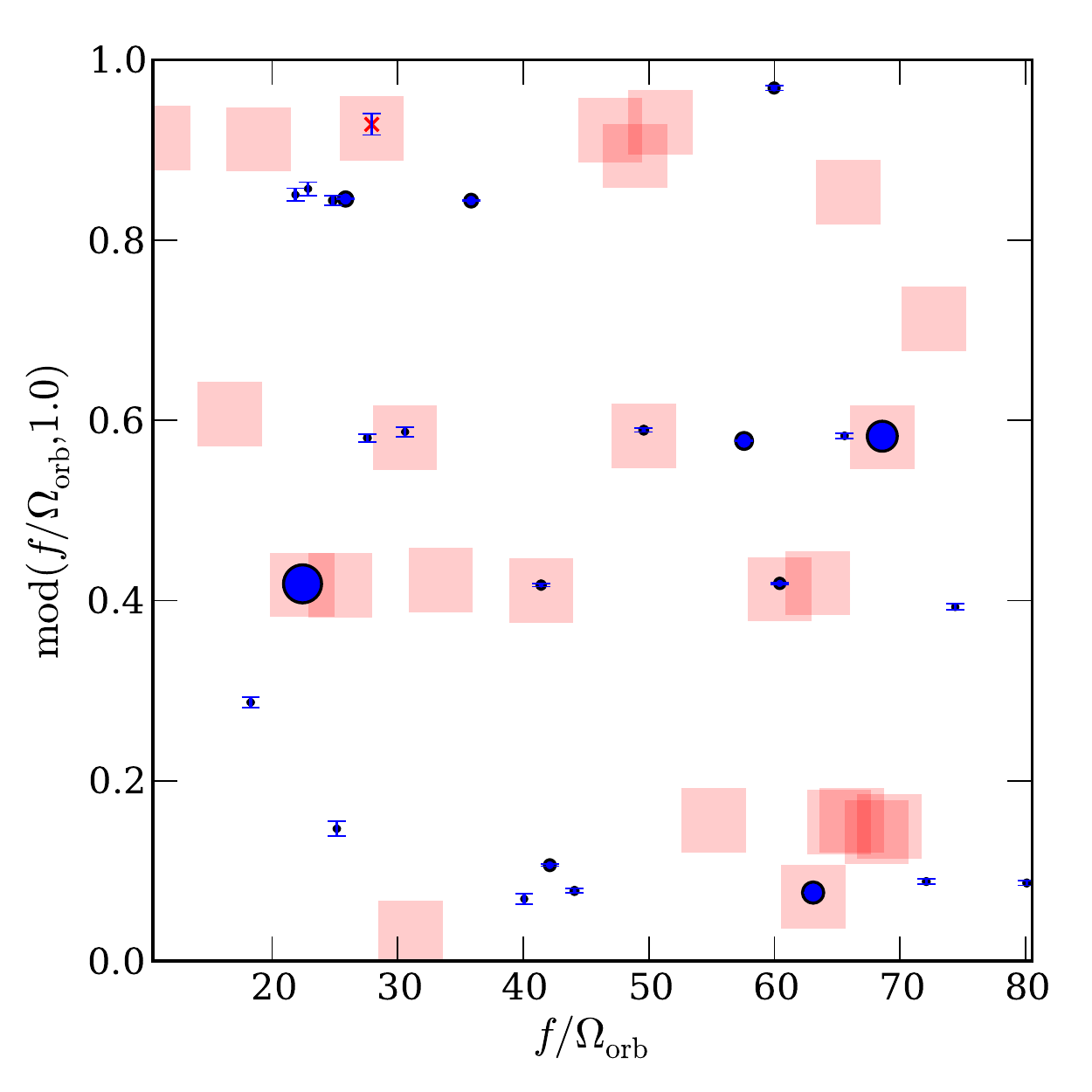}
\caption{Distribution of anharmonic pulsation frequencies.  Plotted is
  the fractional part of $f/\omorb$ as a function of $f/\omorb$ for
  the anharmonic pulsations of KOI-54.  The area of each circle
  corresponds to the amplitude of the pulsation. Assuming that the
  nonlinear coupling is with the 91st harmonic, the complementary
  daughter modes to the anharmonic pulsations are shown as pink
  squares, where $f_{\rm complement}\equiv 91 \omorb - f$.  Four pairs
  of the observed anharmonic pulsations add to $91 \omorb$ within
  their uncertainties: F5 \& F6, F8 \& F100, F25 \& 76, and F39 \&
  F42.  The other pulsations either have complementary pairs that are
  not visible, or are the daughter modes of other pulsations.  This
  figure only shows the pulsations that have frequencies $f/\omorb >
  20$ and amplitudes  $> 1.0\,\mu$mag.}
\label{fig:anharmonic}
\end{figure}

At lower frequencies, the group travel timescale of the wave becomes
comparable to the damping timescale near the surface at the outer
turning point of the mode.  This results in a phase offset of the wave
of order $\sim \sqrt{t_{\rm group} / t_{\rm damp}}$.  As such,
pulsation phases need not obey the relation in
equation~\ref{eq:phase}.  B12 estimated that this should occur for $k
\lesssim 50$ for stars similar to KOI-54 when $m = 0$. These estimates were made
under the approximation of solid-body rotation. For differentially
rotating bodies, the shear between layers could dramatically impact
the onset of the traveling wave regime, resulting in much larger phase
offsets for frequencies $\omega \lesssim 2 \Omega_*$
\citep{1989ApJ...342.1079G}. Only rapidly rotating stars, with much faster spins than expected for KOI-54 (B12), would impact the phases for frequencies much larger than $50\omorb$.

The horizontal dashed line in Figure~\ref{fig:scatterphase} marks
where $k = 50$, approximately when the onset of the traveling regime
is expected (B12).  We see that below this frequency there are more
observed pulsations with phases that are not consistent with $m=0$
standing waves. The phases of the two large-amplitude harmonics,
$k=44$ (F3) and $k=40$ (F4), in particular, are significantly offset
from the phases of the $k=90$ and $k=91$ pulsations. The phase offsets
of traveling waves in KOI-54 appear comparable to the phase offsets in
the best-fit semi-quantitative model of B12 (right panel of
Fig.~\ref{fig:scatterphase}).  For modes with $m \ne 0$, the traveling
wave regime is determined by the frequency of the tide in the
corotating frame, which is Doppler shifted from the observers frame by
$k \omorb - m \Omega_*$.  For $m > 0$, modes with $k\gtrsim 50$ can
still be in the traveling wave regime, especially when it is rotating
rapidly.  In principle, the rotation frequency of the star may be inferred by identifying where
  the $m > 0$ modes enter the traveling wave regime.

We note again that as the pulsation amplitude decreases, it is more
likely that modes from both stars will have comparable contributions
to the pulsations.  This can occur at any frequency, but is more
likely at lower frequencies where the spacing between eigenfrequencies
is smaller (eq.~\ref{eq:asymp}).  When this happens, the observed
phase will be determined by a combination of the pulsation amplitudes
and phases.

\subsection{Time variation of the pulsations}
\label{sec:time}

The estimated timescale for KOI-54 to circularize is very long, $\sim
10^{10}\,$yr, compared to the duration of the observations, so we do
not expect variations in the orbital elements to be directly
observable with the current data set. However, the amplitude and
phases of the pulsations in KOI-54 can vary on much shorter
timescales, especially if the pulsations are in a nonlinear limit
cycle \citep{2001ApJ...546..469W} or are close to resonance ($\delta
\omega \sim \gamma$).  We therefore search for time variation in
the amplitudes, frequencies, and phases of the pulsations.  To do so we
split the data into 6 bins each with equal duration of $131\,$d,
and refit each with the $60$ largest-amplitude pulsations.  We omit 7
pulsations from these fits because their frequencies are spaced too closely together to resolve in the shorter duration of the observations.  We measure the uncertainty in each fit in the same way as for the overall uncertainties: via a Monte
Carlo method with a $1/f$ plus white noise spectrum (see
\S~\ref{sec:osc}).

Figure~\ref{fig:ampchange} shows how the amplitudes of the 90th and
91st harmonic pulsations vary with time. We find that their amplitudes
declined by $3$ and $2$ percent, respectively, over the duration of
the observations.  The characteristic timescales associated with the
changes are $A/{\dot A} \approx 60$ and $90$ years for the 90th and 91st harmonics, respectively.  If
the entire change in amplitude was systematic, we would expect the
change to be the same in both pulsations, but they are different at
the 3$\sigma$ level. We find similar variations in the amplitude of
the two pulsations when using only four or five bins, or when looking
at individual quarters of the \emph{Kepler} data.

To check that our estimated errors are correct, we compare the
distribution of $\delta A/A$ to the estimated error for all of the
observed pulsations.  Excluding the outliers beyond four standard
deviations, we find the estimated errors are only 25 percent smaller than
the variance in $\delta A/A$.  Finally, if the changes are indeed
systematic, we would expect the brightening events at pericenter to
show a similar evolution.  A $3\,$percent error in the absolute
calibration of the source would reduce the peak amplitude of the
tidally induced static tide and irradiation by $\approx 240 \,\mu$mag
over the observed duration of the brightening events.  We do not see
signatures of this deviation in the residuals near the brightening
events even if we only subtract the 20 brightest pulsations.

Such a large change in the amplitude of the pulsations is not
expected. The timescale for the changes is comparable to the damping
time for $n=14$ g-modes.  Even if the oscillations are close to
resonance, which we have already argued the $k=90$ and $k=91$ modes
are likely not, the maximum change in amplitude expected is of order
\begin{equation}
\label{eq:da}
\frac{\delta A}{A} \sim \frac{\Delta t}{t_{\rm tidal}} \frac{\omega_{nl}}{\gamma_{nl}},
\end{equation}
where $\Delta t$ is the duration of the observations, $t_{\rm tidal}$
is the characteristic timescale for tides to change the orbital
parameters, $\gamma_{nl} \approx 1 (n)^4\,$Myr$^{-1}$, with $n = 14$ for $k
\approx 90$ in KOI-54 (W11, B12).  Alternatively we can invert
equation~\ref{eq:da} to get $t_{\rm tidal}$, the timescale of the change in
the orbital parameters.  For $n= 14$ and $ \gamma \approx 4 \times
10^{-2}\,$yr$^{-1}$, $\Delta t \approx 3\,$yr, we derive $t_{\rm tidal}
\lesssim 10^{6}\,$yr. This is a strong upper limit, and is shorter than
the expected synchronization timescale by a factor of $\sim 80$ (B12).

As we alluded to earlier, one possible explanation of our observations
is that the modes in question are undergoing limit cycles as a result
of parametric decay (\S~\ref{sec:nonlin}). This is expected to occur
when the frequency offsets of the daughter modes are less than their
damping rates \citep{2001ApJ...546..469W}. The timescale associated
with the limit cycle, which is $\sim \gamma^{-1}$, is comparable to the
inferred timescale for the amplitude changes.  

However, in \S~\ref{sec:threemode}, we find no evidence to suggest
that the $k=90$ pulsation is undergoing parametric decay. However,
nonlinear second-order coupling between groups of oscillators can
result in chaotic behavior in many circumstances. This can occur even
when the oscillators have the same resonant frequency
\citep[e.g.,][]{chimera}.  

More data or a more sophisticated detrending routine that explicitly
preserves long term trends may be able to determine whether some the
variation of the pulsation amplitudes is systematic or if the
variations are intrinsic to KOI-54. Systematic variations of
approximately one percent have been observed in the transit depths of
the Hot Jupiter HAT-P-7b \citep{2013arXiv1307.6959V}.  The systematic
effects were found to be seasonal, relating to the rotation of the
{\em Kepler} spacecraft. However, \citet{2013arXiv1307.6959V} found no
evidence of systematic trends when comparing data between the same
season.  We find that the amplitudes of the 90th and 91st harmonic
pulsations decreased with time, even when compared to the same season
of data.  Finally, we note that nodal regression owing to a third body
could change the amplitudes of $m\ne 0$ pulsations by altering the
inclination of the system. This would additionally cause changes to
the amplitude of the equilibrium tide, which we do not observe.

\section{Nonlinearly excited tides}
\label{sec:nonlin}
\subsection{Introduction}
\label{sec:nonlinintro}
If a mode achieves a large enough amplitude, the linearized fluid
equations are no longer sufficient to describe its evolution. Instead,
it may decay into daughter modes via the parametric instability.
Traditionally, this has been calculated as a form of resonant
three-mode coupling that minimizes the threshold for the instability
to develop (e.g., \citealt{1982AcA....32..147D},
\citealt{2001ApJ...546..469W}, B12,
\citealt{2011arXiv1107.4594F}). These daughter modes increase the rate
at which energy is damped in the star, potentially accelerating the
impact of the tides. KOI-54 exhibits a variety of pulsations that have
frequencies which are not integer harmonics of the orbital period. In
\S\ref{sec:threemode}, we will explore the distribution of anharmonic
pulsations in KOI-54 in the context of isolated three-mode coupling.
We will show in \S~\ref{sec:multimode} that these anharmonic
pulsations are best explained as the nonlinearly excited daughter
modes of several different parent modes.

Modes can efficiently couple to each other in the second order
when they are near resonance, i.e., when
\begin{equation}
\label{eq:restrict0}
|\omega_A-(\omega_a+\omega_b)| = |\delta \omega_{Aab}| \ll |\omega_A|, 
\end{equation}
where we denote the primary parent mode with the capital subscript $A$
and the the pair of daughter modes as lowercase $a$ and $b$.  They must also satisfy certain
selection rules\footnote{In this work,  we assume that all the frequencies are positive. In B12 the daughter frequencies are considered to be negative, such that $\omega_A+\omega_a+\omega_b = \delta_{Aab}$.  In this convention, equation~\ref{eq:restrict2} would instead read $m_A+m_a+m_b = 0$.}:
\begin{equation}
\label{eq:restrict2}
m_A = m_a+m_b ,
\end{equation}
\begin{equation}
\label{eq:restrict1}
\mod (l_A + l_a + l_b, 2) = 0,
\end{equation}
and
\begin{equation}
\label{eq:restrict3}
|l_a-l_b| \le l_{A} < l_a+l_b.
\end{equation}

Not only
do the intrinsic frequencies of the excited oscillations sum to the parent frequency, but
equation~\ref{eq:restrict2} also implies that the frequencies of
the observed daughter pulsations also sum to the parent frequency: 
\begin{equation}
\label{eq:obsthree}
f_a+f_b = f_A
\end{equation}
For sufficiently small-amplitude oscillations, the daughter modes do
not grow, and the instability is suppressed.  The daughter modes will
only grow in amplitude when the parent mode is sufficiently excited
for the parametric instability to develop (see, e.g., B12).

All modes with $l_a=l_b$ have nonzero coupling coefficients.  However,
for KOI-54 the coupling becomes much less efficient when $l_{a,b} \gg
2$ (B12), so we restrict our attention to $l_{a,b}\le 3$.  For parent
modes with $m_A = 0$, equation~\ref{eq:restrict2} implies $m_a = -
m_b$. Although the onset of parametric decay occurs at relatively
large amplitudes, the complete triplet of modes should not necessarily
be visible in the {\em Kepler} observations. In many cases the parent
mode or one of the daughter modes can have an amplitude that is much
smaller than any of the other oscillations (see, e.g, B12).
Additionally, because KOI-54 is nearly face on, the visibility of $m \ne
0 $ oscillations is suppressed by $\sin^{|m|}{i} \approx (1/10)^{|m|}$.

\begin{figure}
\centering  \includegraphics[width=7cm]{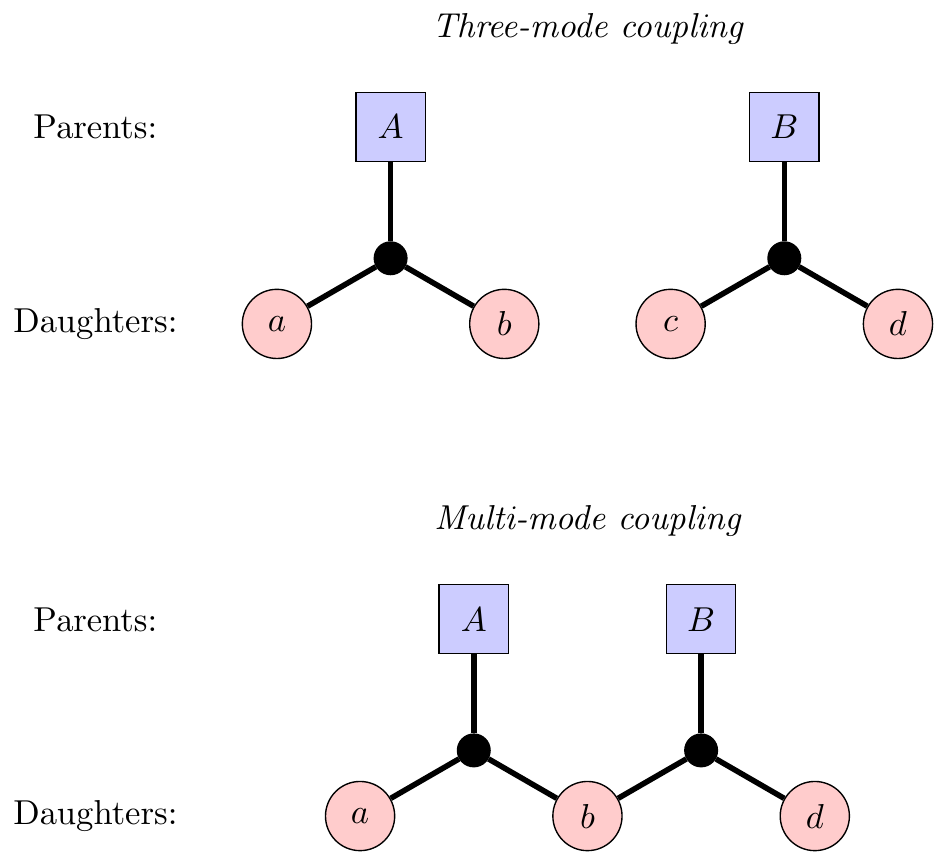}
\caption{Diagram of traditional, isolated three-mode coupling versus
  multiple-mode coupling. The {\em top} panel shows two sets of three
  coupled modes.  In previous nonlinear analyses, these sets are
  treated independently of the other. The {\em bottom} panel shows a
  coupled system of five modes.  It is the same as the top panel---both
  are coupled at second-order---except the two parent modes share the
  coupled daughter mode $b$. KOI-54 shows evidence for at least three
  parent modes coupled to each daughter mode. Because the coupling
  still occurs at second-order, the selection rules for strong
  coupling still apply to multiple-mode coupling. Namely, for the {\em
    bottom} panel, $ f_a+f_b = f_A$, $m_a+m_b = m_A$, and
  additionally $f_b+f_d = f_B$ and $m_b+m_d = m_B$. \label{fig:modecouple}}
\end{figure}
 
\subsection{Three-mode coupling and the observed distribution of daughter pairs}
\label{sec:threemode}
The anharmonic pulsations in KOI-54 are best explained as daughter
modes that are nonlinearly excited by higher-frequency tidally driven
oscillations. As expected from equation~\ref{eq:obsthree}, many of the
anharmonic pulsations have frequencies that, when summed, equal the
frequencies of some of the largest-amplitude harmonic pulsations
within the formal precision of the measurements.  The two most
prominent are the F5 ($f_5 = 22.419\omorb$) and F6 ($f_6 =
68.582\omorb$) pulsations, which appear to be daughter modes of the
91st harmonic of the orbital period. These were the only two
anharmonic pulsations reported in W11 that were consistent with being
a complete daughter mode pair of the $91$st harmonic (B12,
\citealt[]{2011arXiv1107.4594F}).  Here we report a total of four
 pairs of anharmonic pulsations which have frequencies that all add to $91\omorb$.  We
additionally find that the frequencies of many anharmonic pulsations
sum to other harmonic parents, specifically $f_7 = 72\omorb$ and $f_{11}  = 53\omorb$.
These three parent modes in particular have frequencies that are
equally spaced with $\Delta f/\omorb = \Delta k = 19$. We address this in more detail in \S~\ref{sec:multimode}.

In Figure~\ref{fig:anharmonic}, we plot the frequency ($f / \omorb$) and
fractional part of the frequency ($f/\omorb\mod {1.0}$) for each
anharmonic pulsation with measured amplitude $> 1.2\,\mu$mag.  The
area of each circle is proportional to the amplitude of the pulsation.
For each point in the graph, we also highlight the complementary
``sister'' frequency of the pulsation with a pink square, assuming that it
is a daughter of the $k = 91$ harmonic.  In other words, the pink squares represent frequencies that satisfy $f_{\rm
  complement}\equiv 91 \omorb - f$. There are eight points that are
highlighted in pink, corresponding to four complete daughter pairs of
the 91st harmonic. The vast majority of the observed anharmonic
pulsations, however, do not have complementary pairs with amplitudes
above $1.2\,\mu$mag. As discussed in \S~\ref{sec:nonlinintro}, this is
consistent with theoretical expectations.  No two observed anharmonic
pulsations add to $k = 90$.

There are two striking features of the distribution of frequencies
plotted in Figure~\ref{fig:anharmonic}.  First, many of the pulsations
have regular frequency spacing. For all the completely observed
daughter pulsation pairs, there are other observed anharmonic
pulsations with frequencies separated by $\Delta f = \Delta k \omorb =
19 \omorb$.  For example, $f_5+f_6 = f_1 = 91\omorb$, but another
complementary mode sums to the fifth largest-amplitude harmonic
pulsation, $f_5+f_{42} = (91-19)\omorb = f_{7} = 72\omorb$, and to the
sixth largest-amplitude harmonic pulsation $f_5+f_{76} = f_{11} = (91
- 2\times 19)\omorb= 53\omorb$.  A similar pattern emerges for the
daughter pair of F8 and F100: $f_8+f_{100} = f_1$, $f_{49} + f_{100} =
f_7$ and $f_{47}+f_{100} = f_{11}$.  Although it is possible that this
is just a coincidence, we are compelled to suggest that all of these
nonlinearly driven daughter pairs share multiple parent modes. In the
case of KOI-54, three of most prominent parent modes are separated by
$\Delta f = 19 \omorb$. This particular value of $19$ is probably set
randomly by the eigenfrequency spacing unique to one of the stars.
Additionally, each parent mode clearly has multiple daughter pairs. As
we will discuss in \S~\ref{sec:multimode}, the coupling between
multiple parents and daughters is one possible explanation for why the
amplitude of the 91st harmonic is so much lower than the threshold
estimated using only isolated three-mode coupling (B12).
                          
The second salient feature of the anharmonic pulsations is that many
share common fractional (noninteger) parts in units of the orbital
frequency, as was first noted by W11.  There are eight pulsations that have frequencies with
fractional parts near $0.42$  (F5,
F25, F39, F80, F87, F110, F124, \& F128) , and nine observed pulsations  that have frequencies with
their fraction part $0.58$ (F6, F9,
F42, F61, F63, F76, F102, F123, \& F130). That is, if you sum the frequencies
of one pulsation from each group you will get a harmonic (integer)
frequency.  There are also six pulsations  with the fractional part of their frequency near $0.08$ (F8, F47, F49, F58, F59, \&
F75).  There are
eight  near $0.84$ (F17, F19, F48, F71, F72, F86, F108, \& F125).  What
would cause all the pulsations to share similar fractional parts?
Again, we are led to the situation where each daughter is being
nonlinearly excited by multiple parents.  We explore this further in
\S~\ref{sec:multimode}.

The anharmonic pulsations can also be susceptible to the parametric
instability. In this scenario, the daughter modes will themselves couple to 
granddaughter modes, which must obey the same restrictions
(eqs.\ \ref{eq:restrict2}-\ref{eq:restrict3}).  We do not find evidence
for granddaughter modes in KOI-54. However, it would be more difficult to
observe daughters of the already low-frequency anharmonic pulsations in KOI-54.  The
frequency of the largest-amplitude anharmonic pulsation is close to
the frequency cutoff we have imposed in our search for pulsations
($\approx 20\omorb$). Only a pulsation with amplitude $\gtrsim
10\,\mu$mag would be observable above all the noise at low
frequencies.

\subsection{Multiple-mode coupling in KOI-54}
\label{sec:multimode}

\begin{figure*}
\centering \includegraphics[width=\columnwidth]{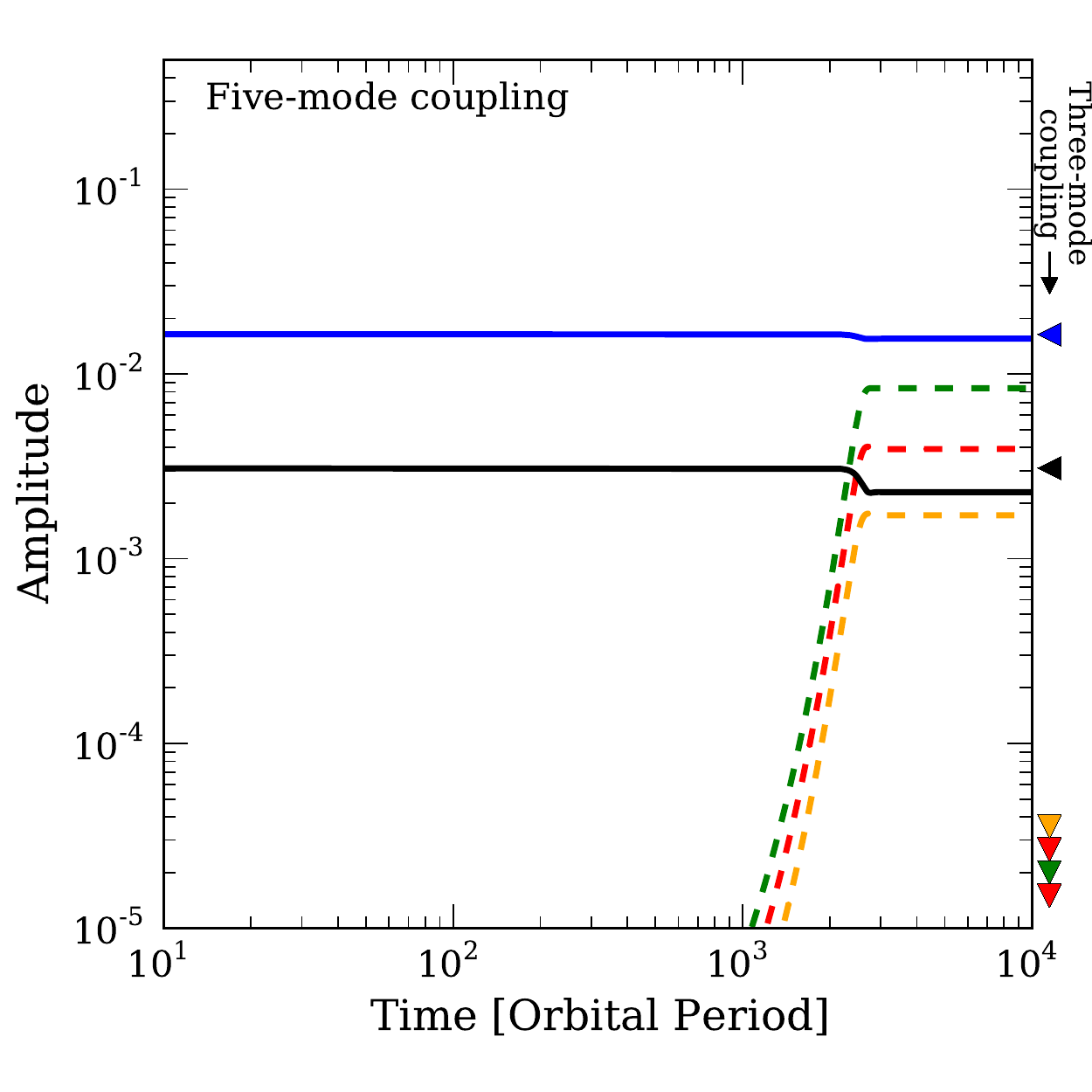}
\includegraphics[width=\columnwidth]{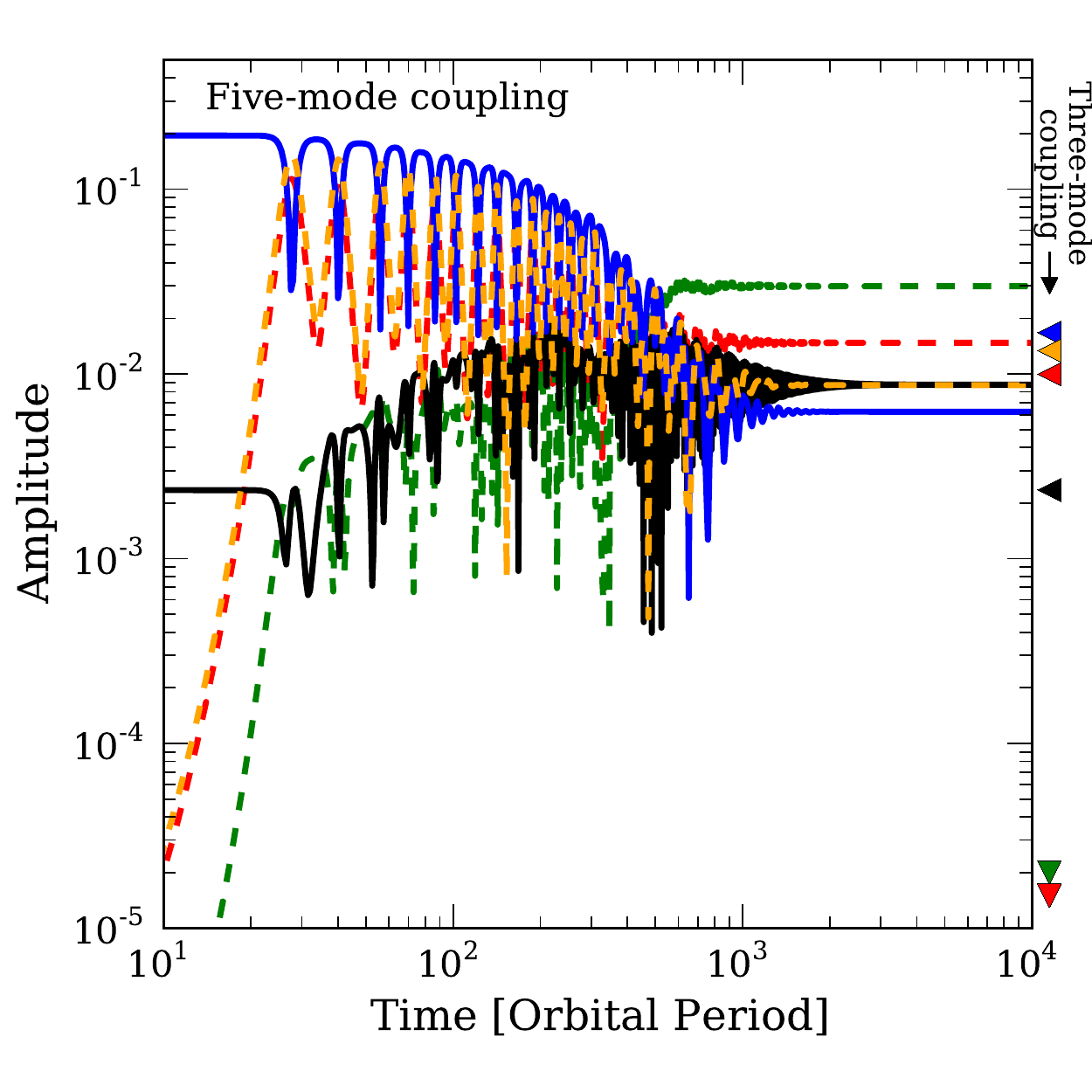}
\caption{The onset of the parametric instability for a network of five
  oscillators coupled at second-order.  In both panels the solid lines
  show the two parent modes, while the dashed lines show the three
  daughter modes for a system of five coupled oscillators. The
  triangles on the right show the final amplitude of the oscillations
  when we integrated the equations for two independent sets of
  three-mode coupling (as shown the top diagram of
  Fig.~\ref{fig:modecouple}). The downward facing triangles indicate
  that the daughter modes did not grow in amplitude.  In the {\em
    left} panel, the system undergoes the parametric instability only
  when treated as five coupled oscillators.  The threshold for the
  onset of the parametric instability is decreased by approximately a
  factor of two compared to isolated three-mode coupling.  The {\em
    right} panel shows how multiple mode coupling can redistribute the
  energy among the different excited modes.  In this case, the largest
  amplitude parent mode is not the one that underwent parametric
  decay, but rather was powered by nonlinear interactions.  We note
  that we did not account for the parametric instability of the
  daughter modes, and so these solutions are not complete, rather they
  are representative of the complex coupling in a network of
  oscillators in a tidally excited star.
\label{fig:pickle}}
\end{figure*}

As described in \S~\ref{sec:threemode}, we find that the anharmonic
pulsations have two features that are not explained in the traditional
picture of three-mode coupling between individual overstable parent
modes and their daughter modes. First, many anharmonic pulsations have
frequency offsets from a perfect harmonic (residual fractions) that
are nearly identical, and second, many observed pairs have common
frequency differences of $\Delta f = 19 \omorb$.  We propose that the
distribution of frequencies in the anharmonic pulsations is naturally
explained by coupling between {\em groups} of daughter pairs and
parents.  Figure~\ref{fig:modecouple} contrasts the case of isolated three-mode
coupling, as we described in \S~\ref{sec:nonlinintro}, and
multiple-mode coupling, as we describe further here.\footnote{Note
  that the multiple-mode coupling scenario we are describing is still
  second-order perturbation theory, and involves only three-mode
  coupling coefficients (see~Fig.~\ref{fig:modecouple}).}  Throughout this section, we use the same terminology as B12.

A common simplifying assumption employed in deriving the nonlinear
parametric instability is to consider coupling only within isolated
mode triplets.  We show this in the top diagram of
Figure~\ref{fig:modecouple}.  By minimizing over all possible sets of
three modes, constrained by equations~\ref{eq:restrict2} --
\ref{eq:restrict3}, one can estimate when the parametric instability
should develop (e.g., \citealt{1982AcA....32..147D},
\citealt{2001ApJ...546..469W}). Recently,
\citet{2012ApJ...751..136W} extended the traditional isolated three-mode
coupling calculations as described in \S~\ref{sec:nonlinintro} to
include the case of a single parent with $N$ distinct daughter pairs.
The authors found this can reduce the threshold for the parent to
become unstable by $\approx 1/N$.  More generally, however, each
eigenmode of the star can additionally couple to multiple parent
modes in second-order perturbation theory, as we show in the bottom diagram of
Figure~\ref{fig:modecouple}.  We explore this scenario by analyzing
the system of equations for five coupled oscillators: two linearly
driven parent modes and two daughter pairs that share a common
eigenmode.  We ask 1) does multiple-mode coupling result in the same
quantitative peculiarities that we observed in \S~\ref{sec:threemode}?
and 2) does multiple-mode coupling lower the threshold for the
parametric instability to develop?

Under the assumption of an equilibrium solution, i.e., that each
eigenmode is a sinusoid with a fixed frequency $f_i$ and amplitude
$A_i$, the observed frequencies will naturally reproduce the trends in
the system.
For a network of five oscillators coupled at second-order, the frequencies of the daughter pairs obey
the relations\footnote{The derivation is similar to the calculation of the equilibrium solution to three-mode coupling in  Appendix D of \citet{2012ApJ...751..136W}.} $f_a + f_b = f_A$ and $f_b+f_d = f_B$ in
order for the presumptive sinusoidal time dependence to cancel out.  If the daughters are coupled to linearly driven
parents, then both $f_A$ and $f_B$ will be harmonics of the orbital
frequency, $k_{A,B} \omorb$.  It then holds that $f_a-f_d = f_B - f_A$
must also be an integer multiple of $\omorb$. Thus, nonlinear
multiple-mode coupling in eccentric binaries manifests itself with
harmonic frequency differences between the anharmonic pulsations, just
as we observe in KOI-54. This property is specific to 
   dynamically excited tides in eccentric binaries. Parent
  oscillations that are driven by instabilities  will not have
  integer spacing.  As such, it becomes less likely that a daughter
  mode will couple to many parents.

Although it is a simple matter to derive the relationship between the
frequencies of a network of five oscillators that are coupled to
second-order, it is not clear if there is a closed-form analytic
equilibrium solution for a network of five (or more) coupled
oscillators, as the steady-state solution of three-mode coupling is
already nontrivial. Instead, we integrate the coupled equations
numerically to look at the consequences of multi-mode coupling on the
steady state properties of the pulsations. In this calculation, we use
realistic damping and driving rates as well as coupling coefficients
computed in B12. The two parents are linearly driven by the tidal
potential at $91\omorb$ and $72\omorb$. 
In this calculation, we use
realistic damping and driving rates as well as coupling coefficients
computed in B12. The two parents are linearly driven by the tidal
potential at $91\omorb$ and $72\omorb$. Rather than trying to represent KOI-54 exactly, we arbitrarily choose the frequencies of the parents and daughter pairs to be close to resonance to demonstrate the impact of multi-mode coupling.  We show the results of
two calculations in Figure~\ref{fig:pickle}. In the two simulations, the intrinsic parent mode frequencies are 
($91.295\omorb$, $72.044\omorb$) and 
($90.615\omorb$, $71.996\omorb$).  In one case, shown in
the left panel, the system undergoes the parametric instability {\em
  only} when the set of five coupled oscillators are solved. When we
treat the oscillators as two independent triplets, the parents are
stable, and no daughter modes are excited. In the right panel, we show
a case where the amplitude of the parent that undergoes the parametric
instability actually ends up pumping energy into an initially
low-amplitude linearly driven oscillation (which was stable in the
isolated three-mode calculations).

To summarize, some consequences of nonlinear multi-mode coupling are:
1) the frequencies of many anharmonic pulsations have the same
residual fraction in units of the orbital frequency; 2) many
anharmonic pulsations may share the same frequency separation; 3) the
largest-amplitude harmonic pulsation is not necessarily the pulsation
undergoing decay (i.e., the energy can be transported via daughter
pairs to another linearly driven parent.); 4) if one parent $l=2$,
$m=0$ is overstable, then all $l=2$, $m=0$ oscillations have some
growth from nonlinear interactions even if all of their amplitudes are
below the threshold for parametric decay.  Furthermore, networks of
coupled oscillators are more likely to enter limit cycles, where the
modes do not reach a steady state, but rather transfer energy back and
forth, sometimes chaotically \citep[see,
  e.g.,][]{2005PhRvD..71f4029B}.

Since the nonlinear coupling is constrained by
equations~\ref{eq:restrict2}-\ref{eq:restrict3}, we can use the
observed anharmonic frequencies to group the pulsations into their
respective modes by degree $l$ and azimuthal order $m$.  For example,
we have identified the 91st harmonic as an $l=2, m=0$ oscillation
(\S~\ref{sec:harm}).  Hence, the daughters pairs that couple to the
91st harmonic in KOI-54 must have $m_a=-m_b$, and most likely have
$l_{a,b} \le 3$. Those pulsations with frequencies that have
noninteger parts near $0.42\omorb$ (e.g., F5) most likely have the
same degree, and must have the opposite azimuthal order of those with
noninteger parts near $0.58\omorb$ (such as F6) since they have
similar observed amplitudes.  The pulsations with noninteger parts
near $0.08$ (F8) must belong to a different chain of non-linear interactions from F5.  Because the
F5 and F6 pulsations are so prominent, we can speculate that these two
pulsations have the smallest degree, $l=1$ or $l=2$, as low degree
pulsations are more easily observable.  There are very few pulsations
observed with noninteger parts near the ``sister'' pair to F8,
F100. One possible reason  is because they are $l=3$ pulsations,
which have higher damping rates and have lower observed amplitudes
since they are averaged out over the disk of the star.

The complicated nature of a network of coupled oscillators makes analyzing
nonlinear systems such as KOI-54 challenging. 
In particular, if there was a mode creating a resonance lock, then it
may only be visible through the nonlinear interactions it has with its
daughter modes, because its observed amplitude may be reduced by
projection effects (\S~\ref{sec:linintro}) and nonlinear interactions
(\S~\ref{sec:nonlinintro}). The two anharmonic pulsations with
frequencies larger than $91\omorb$, F51 and F97, are likely further
evidence of multiple mode-coupling. In isolated three-mode coupling
the resonant parametric instability can only develop for daughter
modes with frequencies less than the parent mode.

\subsection{Additional nonlinear effects}

In linear theory, the observed frequency of a tidally driven
oscillation is always  a perfect harmonic of the orbital period.
However, nonlinear effects can cause the observed frequency of the
oscillation to deviate from a perfect harmonic.  Many of the pulsation
frequencies in Tables~\ref{tab:fitsh} \& \ref{tab:fitsa} have
uncertainties smaller than one part per million.  Although we have not
determined the orbital frequency with a comparable level of precision,
we can still compare two harmonic pulsations to each other and
assess whether  their frequencies deviate from perfect harmonics.  For
the two largest-amplitude pulsations, we see that $f_{91}-91/90
f_{90}\approx 3.31 \pm 0.06\times 10^{-5}\,$day$^{-1}$. This means
that they are inconsistent with both being perfect harmonics of the
orbital frequency by $\sim 50 \sigma$.  Similarly, the 76th and 57th
harmonics deviate from perfect harmonics with respect to the $90$th
harmonic by $\sim 15\sigma$ ($10\,$ppm) and $3\sigma$ ($7\,$ppm) respectively. These deviations are
further evidence of nonlinear interactions in KOI-54.

Multiple-mode coupling may also be an important factor in
some of the observed amplitudes and phases of the harmonic pulsations.
For example, at harmonics that lie between two eigenmodes, $|\delta
\omega|$ is at a maximum, which means that the linear prediction for
the amplitude of the corresponding tidally driven pulsation is
small (see Fig 9. of B12).  In this case, the second-order coupling
between such an oscillation and other tidally driven oscillations (as
well as nonlinearly driven oscillations) may be its dominant source of
energy. That is, second-order effects are likely important for all
oscillations with low intrinsic amplitudes. This not only impacts the
observed amplitude of the pulsation, but also its phase. 

 In addition,
if a parent mode undergoes a significant amount of nonlinear interaction, with a
nonlinear amplitude threshold much less than the expected linear
amplitude, then the observed phase of the pulsation may not reflect
its intrinsic, linear phase (as we assumed in \S~\ref{sec:harm}).  On
the other hand, if we find that the phase offset is near the expected
linear phase, as with the 90th and 91st harmonic pulsations, then
it is likely that its dominant source of energy is  the linear
tide.  We still cannot be certain that the 91st harmonic is the
dominant parent that is driving all the nonlinear effects in KOI-54.  It
has the largest observed amplitude, but as we see in
Figure~\ref{fig:pickle}, the mode that is undergoing parametric decay
can have an intrinsic amplitude  less than the other pulsations in
the system, especially for large $m$ or $l$ oscillations.

Finally, even higher-order nonlinear effects may be present in
KOI-54. As we saw in the introduction of this section, the
second-order coupling between three modes is restricted. In
\S~\ref{sec:multimode}, we saw how these selection rules combined with
nonlinear multiple-mode couplings naturally led to the frequency
spacing of the anharmonic pulsations.  Likewise, third-order coupling
between four or more modes may also impact the observed frequencies of
the anharmonic pulsations.  For example, there are linear combinations
of three pulsations which add to harmonic pulsations, $f_{17}+f_{6}-f_{5} = f_7$
($25.846+68.582-22.419 = 72.009$), as well as many anharmonic pairs
that have frequencies that add to half harmonics.
Theoretical analysis of higher-order nonlinear mode coupling may
provide deeper insight into these relations. Furthermore, future
observations of other similar eccentric binaries
\citep[e.g.,][]{2012ApJ...753...86T,2013arXiv1306.1819H} may show similar features.

\section{Summary and discussion}
\label{sec:disc}

In this paper we have explored the nature and origin of the harmonic
and anharmonic pulsations of the eccentric binary KOI-54. We used 785
days of nearly continuous {\em Kepler} observations of KOI-54 to find
and report over 120 sinusoidal pulsations with frequencies $f \ge 20
\omorb$ and amplitudes as small as $\sim 1\,\mu$mag.  We then used the
phase offset of each harmonic pulsation with respect to the epoch of
periastron to identify the azimuthal order, $m$, of the the
corresponding modes of oscillation.  We also explored the
properties of the anharmonic pulsations; in particular, we focused on
the unique distribution of the pulsation frequencies in KOI-54.

One open question regarding KOI-54 was the nature of the two most
prominent pulsations that have frequencies that are the 90th and 91st
harmonics of the orbital period.  It had been suggested that the large
amplitudes of these two pulsations may be explained if they were
responsible for a resonance lock between the stars' spins and their orbital
motion (\citealt{2011arXiv1107.4594F}; B12).  In \S~\ref{sec:harm}, we used the phase offset of these
pulsations relative to the epoch of periastron to show that they are
most likely $m=0$ modes.  Such oscillations are axisymmetric and thus
cannot be responsible for a resonance lock.  Furthermore, we found
that most of the high frequency $f \gtrsim 50\,\omorb$ harmonic
pulsations in KOI-54 are $m=0$ oscillations.  This is consistent with the prior theoretical expectation that, since KOI-54 is nearly
face-on, only the largest-amplitude $m=\pm 2$ oscillations are
observable.

Oscillations with frequencies $f\lesssim 50\,\omorb$ have
characteristic damping timescales comparable to their group travel
time.  They are thus expected to be traveling waves rather than
standing waves (B12).  In \S~\ref{sec:travel}, we compared the phase
offset of the low-frequency pulsations in KOI-54 to a theoretical
model derived in B12.  We indeed found that  the observed phase
offsets were consistent with the pulsations being traveling waves,
precisely where B12 predicted the oscillations to transition to the
traveling wave regime. Because differential rotation can dramatically
impact on the onset of the traveling wave regime, we conclude that
the approximation of solid-body rotation made in B12 was correct.

In \S~\ref{sec:time} we systematically
searched for time variations in the pulsations' amplitudes and frequencies.
We found that the amplitudes of the 90th and 91st harmonics have
decreased by $3\,\%$ and $2\,\%$, respectively, over the duration of
the observations, corresponding to timescales of $\approx 60$ and $90$ yr, respectively. This is a much larger change than is expected from
linear tidal theory alone.  The 91st harmonic is clearly nonlinearly
coupled to many daughter modes (\S~\ref{sec:nonlin}), and may be in
a limit cycle \citep{2001ApJ...546..469W}, which can cause such rapid
variation. However, the 90th harmonic shows no evidence for nonlinear
coupling.

KOI-54 has a rich population of anharmonic pulsations, which are
naturally explained as being driven by the nonlinear parametric
instability.  Whenever a parent mode exceeds the amplitude threshold
for the parametric instability to develop, it excites daughter mode
pairs, which have frequencies that sum to the parent mode
frequency. Indeed, we observed four complete daughter pairs of the
91st harmonic and several more for the 72nd and 53rd harmonics
(\S~\ref{sec:threemode}). However, there are no two anharmonic
pulsations with frequencies that sum to the 90th harmonic. This is
consistent with the 90th and 91st harmonic pulsations as originating
in different stars in the binary, and only one star undergoing the
parametric instability, although it is unclear why this would be true
since they have similar observed amplitudes.  

Although it was expected that nonlinear interactions can result in
multiple sets of daughter modes, we found strong evidence that the
daughter modes simultaneously couple to multiple parent modes at
second order (\S~\ref{sec:multimode}). Many of the anharmonic
pulsation frequencies are spaced by harmonics of the orbital
frequency.  We showed that a network of oscillators coupled at the
second-order naturally reproduce the characteristic distribution of
frequencies we observe in KOI-54.

Furthermore, we found through
numerical experiments that coupling among multiple modes can lower the
threshold for the onset of nonlinear interactions, and thereby
accelerate the impact of tides, similar to what happens when a single
parent couples to multiple daughters
\citep[see][]{2012ApJ...751..136W}. Because tidally driven
oscillations are naturally spaced as perfect harmonics of the orbital
frequency, we expect multiple-mode coupling to be more common and more
important in highly eccentric systems, such as KOI-54.

Nonlinear effects in eccentric binaries are expected to play an
important role in redistributing the energy and angular momentum
between the orbit and stars
\citep{1996ApJ...466..946K,1998ApJ...507..938G,2010MNRAS.404.1849B,2012ApJ...751..136W}. Since
we have identified the 90th and 91st harmonic pulsations in KOI-54 as
$m=0$ modes, we would like to estimate how much energy they dissipate
compared to the anharmonic pulsations in the system. Unfortunately, we
do not know the azimuthal order of the anharmonic pulsations, and so
we only know their minimum, projected amplitude, which is $\propto
\sin^{|m|}i \approx (1/10)^{|m|}$. Even with this uncertainty, we can
compare the relative damping rates of the modes if we use the scaling
relation that $\gamma \propto \omega^{-4}$ (confirmed for adiabatic
modes in B12), and we also assume that the observed pulsation
amplitude is a constant times the intrinsic amplitude of the
oscillation.  Using the units of B12, the energy dissipation rate in
each mode is $\propto A^2\gamma \propto A^2\omega^{-4}$.  With these
rather uncertain assumptions, the largest nonlinear mode {\em may}
dissipate $(A_5/A_1)^2 (f_5/f_1)^{-4} \sim 20$ times as much energy as
the largest amplitude linearly excited mode.  Even if the asymptotic
relationship for the damping rate of the modes was much shallower,
e.g., $\gamma \propto \omega^{-2}$, the two modes would have
dissipation rates that were comparable.

We have explored the nature of the observed pulsations in KOI-54
without using well tuned stellar models of the stars themselves.
Instead we focused on using qualitatively similar stellar models first
developed in B12, and the asymptotic relations of high-order g-modes
in order to understand the eigenmodes of KOI-54. We have identified
many to be $m=0$ modes of the system using the phase information of
the pulsations. We have also been able to determine that the two
largest amplitude pulsations at the $91$st and $90$th harmonic of the
orbital frequency come from different stars.  In particular we have
identified eight individual anharmonic pulsations that belongs to the
star with the large-amplitude $m=0$ mode at the 91st harmonic of the
orbital frequency. Most, if not all, of the other anharmonic
pulsations belong to this star as well. In future work, it may be
possible to perform much more precise modeling of KOI-54 using these
constraints, as well as in other identified systems
\citep{2012ApJ...753...86T,2013arXiv1306.1819H} with similar analyses.
 
\section*{Acknowledgments}

We would like to thank K.\ Burns, E.\ Petigura, and E.\ Quataert for
useful discussions. This paper includes data collected by the
\emph{Kepler} mission. Funding for the \emph{Kepler} mission is
provided by the NASA Science Mission directorate. All of the
\emph{Kepler} data presented in this paper were obtained from the
Multimission Archive at the Space Telescope Science Institute
(MAST). STScI is operated by the Association of Universities for
Research in Astronomy, Inc., under NASA contract NAS5-26555. Support
for MAST for non-HST data is provided by the NASA Office of Space
Science via grant NNX09AF08G and by other grants and contracts.
R.O.\ is supported by the National Aeronautics and Space
Administration through Einstein Postdoctoral Fellowship Award Number
PF0-110078 issued by the Chandra X-ray Observatory Center, which is
operated by the Smithsonian Astrophysical Observatory for and on
behalf of the National Aeronautics Space Administration under contract
NAS8-03060. J.B.\ is an NSF Graduate Research Fellow.

\bibliography{p}

\clearpage
\newpage
\begin{table*}
\caption{70 largest-amplitude harmonic pulsations with $f/\omorb >
  20$. We use the same IDs for F1 - F30 as W11. Note that we do not
  observe F23, which appears to be indistinguishable from F8. The
  phases of the pulsations are measured relative to BJD-$2454833$
  ($t=0$), the best-fit epoch of periastron found in this work (BF)
  and the best-fit epoch of periastron found by W11
  (W11)\label{tab:fitsh} }
\begin{tabular}{@{}lcrrcrr@{}}
\hline
ID & Frequency  & $f / \omorb$ & Amplitude &  Phase  & Phase & Phase \\
 &  (day$^{-1}$) & & ($\mu$mag) & $(t=0)$ & (BF) & (W11) \\
\hline
F1 & 2.152855~$\pm$~0.000000 & 90.000 & 294.2~$\pm$~0.2 & 0.2865~$\pm$~0.0001 & 0.739 & 0.752\\
F2 & 2.176809~$\pm$~0.000001 & 91.002 & 227.7~$\pm$~0.2 & 0.8139~$\pm$~0.0001 & 0.747 & 0.759\\
F3 & 1.052508~$\pm$~0.000002 & 44.000 & 95.8~$\pm$~0.2 & 0.9116~$\pm$~0.0004 & 0.667 & 0.673\\
F4 & 0.956821~$\pm$~0.000002 & 40.000 & 82.6~$\pm$~0.2 & 0.6626~$\pm$~0.0004 & 0.530 & 0.536\\
F7 & 1.722280~$\pm$~0.000006 & 72.000 & 29.7~$\pm$~0.2 & 0.8036~$\pm$~0.0010 & 0.765 & 0.775\\
F10 & 0.669746~$\pm$~0.000021 & 27.999 & 13.0~$\pm$~0.7 & 0.3235~$\pm$~0.0040 & 0.523 & 0.528\\
F11 & 1.267786~$\pm$~0.000009 & 53.000 & 14.4~$\pm$~0.2 & 0.2699~$\pm$~0.0023 & 0.268 & 0.276\\
F12 & 1.124283~$\pm$~0.000011 & 47.001 & 13.4~$\pm$~0.2 & 0.6279~$\pm$~0.0029 & 0.802 & 0.808\\
F13 & 0.932899~$\pm$~0.000013 & 39.000 & 11.2~$\pm$~0.2 & 0.2310~$\pm$~0.0033 & 0.626 & 0.632\\
F14 & 1.435250~$\pm$~0.000051 & 60.001 & 6.8~$\pm$~0.5 & 0.9278~$\pm$~0.0100 & 0.233 & 0.241\\
F15 & 0.885050~$\pm$~0.000016 & 37.000 & 10.3~$\pm$~0.3 & 0.1727~$\pm$~0.0038 & 0.622 & 0.628\\
F16 & 1.698372~$\pm$~0.000010 & 71.000 & 11.0~$\pm$~0.2 & 0.2348~$\pm$~0.0026 & 0.728 & 0.738\\
F18 & 1.817787~$\pm$~0.000012 & 75.993 & 10.4~$\pm$~0.2 & 0.0435~$\pm$~0.0027 & 0.844 & 0.863\\
F20 & 0.645881~$\pm$~0.000021 & 27.001 & 8.4~$\pm$~0.3 & 0.0754~$\pm$~0.0056 & 0.818 & 0.821\\
F21 & 1.028584~$\pm$~0.000018 & 43.000 & 8.5~$\pm$~0.2 & 0.9820~$\pm$~0.0041 & 0.264 & 0.271\\
F22 & 1.076414~$\pm$~0.000019 & 45.000 & 8.8~$\pm$~0.2 & 0.0337~$\pm$~0.0046 & 0.257 & 0.264\\
F24 & 0.861143~$\pm$~0.000028 & 36.000 & 6.3~$\pm$~0.3 & 0.3958~$\pm$~0.0061 & 0.377 & 0.382\\
F26 & 1.243886~$\pm$~0.000021 & 52.001 & 7.1~$\pm$~0.2 & 0.6981~$\pm$~0.0042 & 0.230 & 0.237\\
F28 & 0.789395~$\pm$~0.000032 & 33.001 & 5.7~$\pm$~0.3 & 0.2535~$\pm$~0.0078 & 0.824 & 0.828\\
F29 & 0.693810~$\pm$~0.000047 & 29.005 & 4.4~$\pm$~0.3 & 0.8189~$\pm$~0.0098 & 0.528 & 0.528\\
F30 & 1.148197~$\pm$~0.000024 & 48.000 & 5.9~$\pm$~0.2 & 0.5982~$\pm$~0.0060 & 0.242 & 0.248\\
F32 & 1.865862~$\pm$~0.000028 & 78.002 & 5.1~$\pm$~0.2 & 0.5579~$\pm$~0.0051 & 0.365 & 0.374\\
F33 & 1.172065~$\pm$~0.000029 & 48.998 & 5.1~$\pm$~0.2 & 0.1353~$\pm$~0.0072 & 0.236 & 0.245\\
F34 & 0.765372~$\pm$~0.000042 & 31.996 & 4.7~$\pm$~0.3 & 0.9288~$\pm$~0.0094 & 0.999 & 0.008\\
F35 & 1.363385~$\pm$~0.000030 & 56.996 & 4.5~$\pm$~0.2 & 0.3559~$\pm$~0.0075 & 0.218 & 0.230\\
F36 & 1.100361~$\pm$~0.000036 & 46.001 & 4.3~$\pm$~0.2 & 0.5158~$\pm$~0.0085 & 0.217 & 0.224\\
F37 & 0.741512~$\pm$~0.000046 & 30.999 & 4.2~$\pm$~0.3 & 0.3690~$\pm$~0.0083 & 0.984 & 0.990\\
F38 & 0.621861~$\pm$~0.000056 & 25.997 & 4.1~$\pm$~0.3 & 0.8731~$\pm$~0.0117 & 0.117 & 0.124\\
F40 & 1.004723~$\pm$~0.000046 & 42.002 & 4.0~$\pm$~0.3 & 0.0634~$\pm$~0.0105 & 0.890 & 0.894\\
F41 & 1.219930~$\pm$~0.000035 & 50.999 & 4.0~$\pm$~0.2 & 0.1753~$\pm$~0.0085 & 0.226 & 0.234\\
F43 & 1.315663~$\pm$~0.000042 & 55.001 & 3.6~$\pm$~0.2 & 0.2478~$\pm$~0.0092 & 0.199 & 0.206\\
F44 & 0.837307~$\pm$~0.000054 & 35.004 & 3.4~$\pm$~0.3 & 0.3227~$\pm$~0.0119 & 0.855 & 0.857\\
F45 & 1.195970~$\pm$~0.000040 & 49.998 & 3.4~$\pm$~0.2 & 0.6736~$\pm$~0.0096 & 0.242 & 0.252\\
F46 & 0.598074~$\pm$~0.000238 & 25.003 & 3.0~$\pm$~5.1 & 0.2808~$\pm$~0.0617 & 0.089 & 0.090\\
F50 & 0.908896~$\pm$~0.000050 & 37.996 & 2.9~$\pm$~0.3 & 0.3940~$\pm$~0.0134 & 0.295 & 0.304\\
F52 & 0.526192~$\pm$~0.000079 & 21.997 & 2.8~$\pm$~0.3 & 0.7832~$\pm$~0.0162 & 0.144 & 0.150\\
F53 & 0.813166~$\pm$~0.000066 & 33.994 & 2.6~$\pm$~0.3 & 0.7702~$\pm$~0.0164 & 0.771 & 0.782\\
F54 & 0.717522~$\pm$~0.000073 & 29.996 & 2.5~$\pm$~0.3 & 0.9400~$\pm$~0.0167 & 0.065 & 0.073\\
F55 & 0.574141~$\pm$~0.000100 & 24.002 & 2.5~$\pm$~0.3 & 0.7270~$\pm$~0.0173 & 0.060 & 0.062\\
F56 & 0.549848~$\pm$~0.000227 & 22.986 & 2.1~$\pm$~2.2 & 0.4992~$\pm$~0.0426 & 0.260 & 0.277\\
F57 & 3.037914~$\pm$~0.000040 & 127.000 & 2.1~$\pm$~0.2 & 0.9778~$\pm$~0.0108 & 0.882 & 0.901\\
F60 & 1.291750~$\pm$~0.000069 & 54.002 & 2.0~$\pm$~0.2 & 0.6138~$\pm$~0.0141 & 0.096 & 0.102\\
F62 & 1.339562~$\pm$~0.000082 & 56.000 & 1.9~$\pm$~0.2 & 0.6962~$\pm$~0.0173 & 0.113 & 0.121\\
F64 & 1.674427~$\pm$~0.000078 & 69.999 & 1.8~$\pm$~0.2 & 0.2154~$\pm$~0.0159 & 0.230 & 0.241\\
F65 & 2.057235~$\pm$~0.000077 & 86.003 & 1.7~$\pm$~0.2 & 0.1383~$\pm$~0.0179 & 0.721 & 0.731\\
F67 & 2.511686~$\pm$~0.000069 & 105.001 & 1.6~$\pm$~0.2 & 0.6513~$\pm$~0.0126 & 0.185 & 0.200\\
F68 & 4.090343~$\pm$~0.000054 & 170.997 & 1.6~$\pm$~0.1 & 0.8614~$\pm$~0.0129 & 0.500 & 0.528\\
F69 & 2.009328~$\pm$~0.000078 & 84.000 & 1.6~$\pm$~0.2 & 0.6009~$\pm$~0.0167 & 0.223 & 0.235\\
F73 & 1.459136~$\pm$~0.000091 & 60.999 & 1.5~$\pm$~0.2 & 0.4344~$\pm$~0.0188 & 0.202 & 0.212\\
F74 & 1.387403~$\pm$~0.000089 & 58.000 & 1.5~$\pm$~0.2 & 0.7848~$\pm$~0.0212 & 0.145 & 0.153\\
F77 & 1.626530~$\pm$~0.000101 & 67.997 & 1.5~$\pm$~0.2 & 0.1719~$\pm$~0.0215 & 0.228 & 0.241\\
F78 & 0.980710~$\pm$~0.000106 & 40.999 & 1.5~$\pm$~0.2 & 0.2836~$\pm$~0.0251 & 0.614 & 0.621\\
F79 & 1.842045~$\pm$~0.000083 & 77.007 & 1.4~$\pm$~0.2 & 0.3939~$\pm$~0.0234 & 0.757 & 0.762\\
F82 & 0.502794~$\pm$~0.000304 & 21.019 & 1.4~$\pm$~0.3 & 0.9188~$\pm$~0.0429 & 0.949 & 0.933\\
F83 & 1.530989~$\pm$~0.000096 & 64.003 & 1.4~$\pm$~0.2 & 0.8901~$\pm$~0.0213 & 0.098 & 0.104\\
F85 & 1.650321~$\pm$~0.000098 & 68.992 & 1.4~$\pm$~0.2 & 0.7356~$\pm$~0.0198 & 0.228 & 0.247\\
F89 & 2.368300~$\pm$~0.000088 & 99.007 & 1.2~$\pm$~0.2 & 0.9339~$\pm$~0.0219 & 0.675 & 0.683\\
F90 & 1.483000~$\pm$~0.000120 & 61.997 & 1.2~$\pm$~0.2 & 0.9956~$\pm$~0.0257 & 0.220 & 0.232\\
F92 & 1.602687~$\pm$~0.000136 & 67.000 & 1.2~$\pm$~0.2 & 0.4568~$\pm$~0.0327 & 0.062 & 0.072\\
F93 & 1.411296~$\pm$~0.000130 & 58.999 & 1.2~$\pm$~0.2 & 0.3565~$\pm$~0.0305 & 0.181 & 0.191\\

\end{tabular}
\end{table*}
\begin{table*}
\contcaption{}
\begin{tabular}{@{}lcrrcrr@{}}
\hline
ID & Frequency  & $f / \omorb$ & Amplitude &  Phase  & Phase & Phase \\
 &  (day$^{-1}$) & & ($\mu$mag) & $(t=0)$ & (BF) & (W11) \\
\hline
F98 & 1.793788~$\pm$~0.000108 & 74.989 & 1.1~$\pm$~0.2 & 0.9890~$\pm$~0.0266 & 0.297 & 0.318\\
F101 & 2.105097~$\pm$~0.000110 & 88.004 & 1.1~$\pm$~0.2 & 0.6535~$\pm$~0.0286 & 0.186 & 0.195\\
F112 & 1.578690~$\pm$~0.000171 & 65.997 & 0.9~$\pm$~0.2 & 0.9594~$\pm$~0.0313 & 0.072 & 0.085\\
F114 & 1.937589~$\pm$~0.000153 & 81.001 & 0.9~$\pm$~0.2 & 0.0269~$\pm$~0.0346 & 0.240 & 0.251\\
F115 & 1.985442~$\pm$~0.000147 & 83.001 & 0.9~$\pm$~0.2 & 0.0068~$\pm$~0.0331 & 0.166 & 0.177\\
F116 & 1.746554~$\pm$~0.000170 & 73.015 & 0.9~$\pm$~0.2 & 0.2281~$\pm$~0.0409 & 0.757 & 0.753\\
F117 & 1.555039~$\pm$~0.000166 & 65.008 & 0.9~$\pm$~0.2 & 0.2583~$\pm$~0.0370 & 0.972 & 0.974\\
F119 & 2.248625~$\pm$~0.000154 & 94.004 & 0.8~$\pm$~0.2 & 0.2845~$\pm$~0.0354 & 0.648 & 0.658\\
F127 & 2.080996~$\pm$~0.000189 & 86.996 & 0.7~$\pm$~0.1 & 0.8380~$\pm$~0.0463 & 0.849 & 0.866\\
F129 & 1.889982~$\pm$~0.000210 & 79.011 & 0.7~$\pm$~0.2 & 0.4132~$\pm$~0.0454 & 0.746 & 0.747\\

\end{tabular}
\end{table*}
\clearpage

\begin{table*}
\caption{59 largest-amplitude anharmonic pulsations with $f/\omorb > 20$. We use the same IDs for F1 - F30 as W11. The
  phases of the pulsations are measured relative to BJD-$2454833$
  ($t=0$), the best-fit epoch of periastron found in this work (BF)
  and the best-fit epoch of periastron found by W11
  (W11)\label{tab:fitsa} }
\begin{tabular}{@{}lcrrcrr@{}}
\hline
ID & Frequency  & $f / \omorb$ & Amplitude &  Phase  & Phase & Phase \\
 &  (day$^{-1}$) & & ($\mu$mag) & $(t=0)$ & (BF) & (W11) \\
\hline
F5 & 0.536266~$\pm$~0.000002 & 22.419 & 78.7~$\pm$~0.3 & 0.5181~$\pm$~0.0006 & 0.604 & 0.189\\
F6 & 1.640532~$\pm$~0.000003 & 68.582 & 49.0~$\pm$~0.2 & 0.1187~$\pm$~0.0006 & 0.963 & 0.391\\
F8 & 1.508813~$\pm$~0.000006 & 63.076 & 24.6~$\pm$~0.2 & 0.7570~$\pm$~0.0013 & 0.965 & 0.898\\
F9 & 1.377281~$\pm$~0.000009 & 57.577 & 15.7~$\pm$~0.2 & 0.5055~$\pm$~0.0019 & 0.128 & 0.559\\
F17 & 0.618246~$\pm$~0.000019 & 25.846 & 11.2~$\pm$~0.3 & 0.1672~$\pm$~0.0044 & 0.433 & 0.591\\
F19 & 0.857408~$\pm$~0.000019 & 35.844 & 9.0~$\pm$~0.2 & 0.3692~$\pm$~0.0044 & 0.340 & 0.501\\
F25 & 1.445260~$\pm$~0.000021 & 60.419 & 5.9~$\pm$~0.2 & 0.3859~$\pm$~0.0047 & 0.399 & 0.989\\
F27 & 1.007207~$\pm$~0.000025 & 42.106 & 6.6~$\pm$~0.2 & 0.8551~$\pm$~0.0057 & 0.354 & 0.254\\
F31 & 1.434493~$\pm$~0.000061 & 59.969 & 5.7~$\pm$~0.5 & 0.0866~$\pm$~0.0134 & 0.187 & 0.227\\
F39 & 0.990723~$\pm$~0.000043 & 41.417 & 4.1~$\pm$~0.2 & 0.9145~$\pm$~0.0106 & 0.954 & 0.543\\
F42 & 1.186201~$\pm$~0.000049 & 49.589 & 3.6~$\pm$~0.2 & 0.1480~$\pm$~0.0091 & 0.074 & 0.492\\
F47 & 0.599822~$\pm$~0.000158 & 25.076 & 3.0~$\pm$~1.1 & 0.1305~$\pm$~0.0192 & 0.412 & 0.340\\
F48 & 0.594286~$\pm$~0.000127 & 24.844 & 2.9~$\pm$~0.3 & 0.8995~$\pm$~0.0178 & 0.683 & 0.842\\
F49 & 1.054365~$\pm$~0.000057 & 44.078 & 2.9~$\pm$~0.2 & 0.6138~$\pm$~0.0127 & 0.871 & 0.800\\
F51 & 2.229333~$\pm$~0.000039 & 93.197 & 2.9~$\pm$~0.2 & 0.5309~$\pm$~0.0093 & 0.675 & 0.491\\
F58 & 1.915717~$\pm$~0.000065 & 80.087 & 2.1~$\pm$~0.2 & 0.5274~$\pm$~0.0148 & 0.823 & 0.748\\
F59 & 1.724390~$\pm$~0.000072 & 72.088 & 2.0~$\pm$~0.2 & 0.8363~$\pm$~0.0144 & 0.368 & 0.291\\
F61 & 0.659742~$\pm$~0.000101 & 27.581 & 2.0~$\pm$~0.3 & 0.0812~$\pm$~0.0209 & 0.574 & 0.997\\
F63 & 1.568781~$\pm$~0.000068 & 65.583 & 1.8~$\pm$~0.2 & 0.9627~$\pm$~0.0170 & 0.395 & 0.822\\
F66 & 0.601524~$\pm$~0.000197 & 25.147 & 1.7~$\pm$~0.3 & 0.0434~$\pm$~0.0288 & 0.785 & 0.642\\
F70 & 1.779522~$\pm$~0.000080 & 74.393 & 1.6~$\pm$~0.2 & 0.2958~$\pm$~0.0180 & 0.744 & 0.362\\
F71 & 0.546754~$\pm$~0.000177 & 22.857 & 1.6~$\pm$~0.3 & 0.7310~$\pm$~0.0336 & 0.655 & 0.801\\
F72 & 0.522676~$\pm$~0.000171 & 21.850 & 1.6~$\pm$~0.3 & 0.4212~$\pm$~0.0335 & 0.831 & 0.983\\
F75 & 0.958470~$\pm$~0.000142 & 40.069 & 1.5~$\pm$~0.2 & 0.5441~$\pm$~0.0265 & 0.857 & 0.794\\
F76 & 0.731666~$\pm$~0.000125 & 30.587 & 1.5~$\pm$~0.2 & 0.0506~$\pm$~0.0295 & 0.002 & 0.419\\
F80 & 0.487952~$\pm$~0.000158 & 20.399 & 1.4~$\pm$~0.3 & 0.1995~$\pm$~0.0355 & 0.215 & 0.819\\
F81 & 0.542831~$\pm$~0.000275 & 22.693 & 1.4~$\pm$~8.1 & 0.4005~$\pm$~0.0387 & 0.263 & 0.573\\
F84 & 0.501041~$\pm$~0.000391 & 20.946 & 1.4~$\pm$~1.1 & 0.0086~$\pm$~0.0695 & 0.565 & 0.622\\
F86 & 0.571241~$\pm$~0.000200 & 23.881 & 1.4~$\pm$~0.3 & 0.1972~$\pm$~0.0407 & 0.746 & 0.869\\
F87 & 1.110606~$\pm$~0.000128 & 46.429 & 1.3~$\pm$~0.2 & 0.7899~$\pm$~0.0243 & 0.263 & 0.841\\
F88 & 2.158883~$\pm$~0.000110 & 90.252 & 1.3~$\pm$~0.2 & 0.4442~$\pm$~0.0208 & 0.528 & 0.289\\
F91 & 0.551048~$\pm$~0.000546 & 23.037 & 1.2~$\pm$~9.3 & 0.7859~$\pm$~0.0672 & 0.871 & 0.838\\
F94 & 0.816549~$\pm$~0.000168 & 34.136 & 1.2~$\pm$~0.3 & 0.4332~$\pm$~0.0372 & 0.350 & 0.219\\
F95 & 0.569195~$\pm$~0.000213 & 23.795 & 1.1~$\pm$~0.3 & 0.3106~$\pm$~0.0462 & 0.306 & 0.514\\
F96 & 0.495559~$\pm$~0.000192 & 20.717 & 1.1~$\pm$~0.3 & 0.1880~$\pm$~0.0459 & 0.261 & 0.547\\
F97 & 4.710063~$\pm$~0.000080 & 196.904 & 1.1~$\pm$~0.1 & 0.7782~$\pm$~0.0179 & 0.081 & 0.206\\
F99 & 1.189604~$\pm$~0.000168 & 49.731 & 1.1~$\pm$~0.2 & 0.6696~$\pm$~0.0347 & 0.516 & 0.792\\
F100 & 0.668068~$\pm$~0.000282 & 27.929 & 1.1~$\pm$~0.6 & 0.7573~$\pm$~0.0483 & 0.502 & 0.578\\
F102 & 0.491229~$\pm$~0.000293 & 20.536 & 1.0~$\pm$~0.3 & 0.0509~$\pm$~0.0475 & 0.952 & 0.420\\
F103 & 0.605359~$\pm$~0.000251 & 25.307 & 1.0~$\pm$~0.3 & 0.2140~$\pm$~0.0523 & 0.993 & 0.690\\
F104 & 0.552568~$\pm$~0.000537 & 23.100 & 1.0~$\pm$~9.3 & 0.3554~$\pm$~0.0895 & 0.852 & 0.755\\
F105 & 0.579934~$\pm$~0.000275 & 24.244 & 0.9~$\pm$~0.3 & 0.5667~$\pm$~0.0580 & 0.467 & 0.227\\
F106 & 0.988454~$\pm$~0.000240 & 41.322 & 0.9~$\pm$~0.2 & 0.2843~$\pm$~0.0436 & 0.710 & 0.393\\
F107 & 0.508082~$\pm$~0.000761 & 21.240 & 0.9~$\pm$~0.3 & 0.3580~$\pm$~0.1756 & 0.819 & 0.582\\
F108 & 0.499611~$\pm$~0.000623 & 20.886 & 0.9~$\pm$~1.2 & 0.1867~$\pm$~0.0800 & 0.356 & 0.473\\
F109 & 0.596670~$\pm$~0.000627 & 24.944 & 0.9~$\pm$~5.2 & 0.1192~$\pm$~0.0968 & 0.548 & 0.608\\
F110 & 0.584317~$\pm$~0.000296 & 24.427 & 0.9~$\pm$~0.3 & 0.9406~$\pm$~0.0655 & 0.027 & 0.603\\
F111 & 0.629407~$\pm$~0.000260 & 26.312 & 0.9~$\pm$~0.2 & 0.5700~$\pm$~0.0533 & 0.855 & 0.547\\
F113 & 0.564864~$\pm$~0.000450 & 23.614 & 0.9~$\pm$~1.3 & 0.4037~$\pm$~0.0627 & 0.227 & 0.616\\
F118 & 1.379613~$\pm$~0.000201 & 57.675 & 0.8~$\pm$~0.2 & 0.0054~$\pm$~0.0416 & 0.258 & 0.592\\
F120 & 0.912653~$\pm$~0.000353 & 38.153 & 0.8~$\pm$~0.2 & 0.7466~$\pm$~0.0572 & 0.664 & 0.516\\
F121 & 0.557153~$\pm$~0.000506 & 23.292 & 0.8~$\pm$~0.3 & 0.5440~$\pm$~0.0937 & 0.281 & 0.993\\
F122 & 0.562059~$\pm$~0.000504 & 23.497 & 0.8~$\pm$~1.3 & 0.9654~$\pm$~0.0870 & 0.030 & 0.537\\
F123 & 0.635773~$\pm$~0.000293 & 26.579 & 0.8~$\pm$~0.2 & 0.1484~$\pm$~0.0666 & 0.156 & 0.582\\
F124 & 1.038741~$\pm$~0.000234 & 43.425 & 0.8~$\pm$~0.2 & 0.8243~$\pm$~0.0495 & 0.855 & 0.437\\
F125 & 0.689621~$\pm$~0.000494 & 28.830 & 0.7~$\pm$~0.2 & 0.0524~$\pm$~0.3147 & 0.629 & 0.803\\
F126 & 0.544352~$\pm$~0.001135 & 22.757 & 0.7~$\pm$~8.1 & 0.8131~$\pm$~0.1062 & 0.087 & 0.334\\
F128 & 1.277955~$\pm$~0.000270 & 53.425 & 0.7~$\pm$~0.2 & 0.7726~$\pm$~0.0579 & 0.522 & 0.105\\
F130 & 0.803000~$\pm$~0.000427 & 33.569 & 0.7~$\pm$~0.2 & 0.6504~$\pm$~0.2541 & 0.901 & 0.337\\

\end{tabular}
\end{table*}

\end{document}